\documentclass[12pt]{article}
\usepackage{color}
\usepackage{bm}
\usepackage{graphicx}
\usepackage{amsmath}
\usepackage{amssymb}
\usepackage{fullpage}
\usepackage{hyperref}
\usepackage{float}

\newcommand\pubdate{}

\def\STANKI{Deptartment of Physics and Kavli Institute for Astrophysics and Space Research\\
 Massachusetts Institute of Technology , Cambridge, Massachusetts 02139, USA}

\def\Title#1{\begin{center} {\Large #1 } \end{center}}
\def\Author#1{\begin{center}{ \sc #1} \end{center}}
\def\Address#1{\begin{center}{ \it #1} \end{center}}

\newcommand\pubblock{\rightline{\begin{tabular}{l}\\
         \pubdate \end{tabular}}}

\newcommand{\lcdm}{$\Lambda$CDM }

\newcommand{\bp}{ \bar{\varphi}}

\newcommand{\mpl}{m_{\mbox{\tiny{Pl}}}}
\newcommand{\aosc}{a_{\rm{osc}}}
\newcommand{\Beq}{\begin{equation}\begin{aligned}}
\newcommand{\Eeq}{\end{aligned}\end{equation}}

\newcommand{\vp}{\varphi}
\newcommand{\anl}{a_{\rm{nl}}}
\newcommand{\bx}{{\bf{x}}}
\newcommand{\bn}{{\bf{n}}}
\newcommand{\knl}{k_{\rm{nl}}}
\newcommand{\Rnl}{R_{\rm{nl}}}
\newcommand{\tnl}{t_{\rm{nl}}}

\usepackage{ wasysym }

\date{ }
\begin{document}

\begin{titlepage}
\pubblock

\vfill \Title{Scale-dependent growth from a transition in dark energy dynamics}
\vfill \Author{Mustafa A. Amin\footnote{mamin@mit.edu}, Phillip Zukin\footnote{zukin@mit.edu} and Edmund Bertschinger\footnote{edbert@mit.edu}} \Address{\STANKI} 
\abstract

We investigate the observational consequences of the quintessence field rolling to and oscillating near a minimum in its potential, {\it{if}} it happens close to the present epoch ($z \lesssim 0.2$). We show that in a class of models, the oscillations lead to a rapid growth of the field fluctuations and the gravitational potential on subhorizon scales. The growth in the gravitational potential occurs on timescales $\ll H^{-1}$. This effect is present even when the quintessence parameters are chosen to reproduce an expansion history consistent with observations. For linearized fluctuations, we find that although the gravitational potential power spectrum is enhanced in a scale-dependent manner, the shape of the dark matter/galaxy power spectrum is not significantly affected. We find that the best constraints on such a transition in the quintessence field is provided via the integrated Sachs-Wolfe effect in the CMB temperature power spectrum. Going beyond the linearized regime, the quintessence field can fragment into large, localized, long-lived excitations (oscillons) with sizes comparable to galaxy clusters; this fragmentation could provide additional observational constraints. Two quoted signatures of modified gravity are a scale-dependent  growth of the gravitational potential and a difference between the matter power spectrum inferred from measurements of lensing and galaxy clustering. Here, both effects are achieved by a minimally coupled scalar field in general relativity with a canonical kinetic term. In other words we show that, with some tuning of parameters, scale-dependent growth does not necessarily imply a violation of General Relativity.

\vfill
\end{titlepage}
\def\thefootnote{\fnsymbol{footnote}}
\setcounter{footnote}{0}
\section{Introduction}
Scalar fields are used ubiquitously in cosmology to provide a mechanism for accelerated expansion during both the inflationary \cite{Guth:1980zm,Linde:1981mu,Albrecht:1982wi} and current dark energy dominated epochs (for example, \cite{Ratra:1987rm,ArmendarizPicon:2000dh,Copeland:2006wr}). In the inflationary case, accelerated expansion occurs when the inflaton field is rolling slowly and it ends once the field starts oscillating around the minimum of its potential. The oscillatory phase is well studied and gives rise to rapid growth of inhomogeneities because of couplings to other fields or self interactions (see, e.g. \cite{Traschen:1990sw,Kofman:1994rk, Amin:2010xe,Amin:2010dc}). For a recent review, see \cite{Allahverdi:2010xz}. Like the inflaton, the quintessence field provides a mechanism for accelerated expansion in the slow-roll regime. However, unlike inflation, the possibility of the accelerated expansion ending through quintessence oscillations is rarely discussed in the literature (however, see e.g. \cite{Frieman:1995pm}). If quintessence was rolling slowly in the past, it is ``natural" (though certainly not necessary) for it to enter an oscillatory regime as it finds its way towards a local minimum in its potential. 

In this paper, we wish to understand the observational signatures of a transition to an oscillatory regime in the quintessence field, if such a transition occurs in our recent past ($z\lesssim 0.2$). An obvious observational signature is a late-time change in expansion history. More importantly, as is the case with the inflaton, the coherent oscillations of quintessence near an anharmonic minimum are unstable and quickly fragment in a spatially inhomogeneous manner, giving rise to additional structure. This instability driven by self interactions of the field (the anharmonic terms) leads to a growth in structure that is much faster than  the usual gravitational growth. For example, all potentials that have a quadratic minimum and are shallower than quadratic away from the minimum (see Fig. \ref{fig:potential}) suffer from this instability (in particular for wavenumbers $k\ll m$, where $m^2\approx U''(\vp)|_{\vp\rightarrow0}$. See, for e.g \cite{Khlopov:1985jw,Johnson:2008se,Amin:2010xe, Amin:2011hj}). We find that while the gravitational potential is influenced strongly by such growth in the quintessence fluctuations, the over density in baryons and dark matter does not change significantly. The scale dependent growth in the gravitational potential and the {\it absence} of similar growth in the matter power spectrum provides a signature to confirm or rule out such a transition in the dynamics of the quintessence field. In addition, if the quintessence field becomes nonlinear, it can fragment rapidly into long-lived, localized excitations called oscillons (see e.g. \cite{Bogolyubsky:1976yu, Segur:1987mg, Gleiser:1993pt, Copeland:1995fq, McDonald:2001iv, Kasuya:2002zs,Gleiser:2004iy, Gleiser:2006te, Hindmarsh:2006ur, Saffin:2006yk, Farhi:2007wj, Fodor:2009kf, Gleiser:2009ys, Amin:2010xe, Amin:2010jq, Hertzberg:2010yz, Gleiser:2010qt, Gleiser:2011xj,Amin:2011hj}) with size $l\sim {\rm{few}}\times m^{-1}\sim\rm{few}\times\rm{Mpc}$\footnote{We will frequently relate the mass of the scalar field $m$, to length and time scales, without explicitly writing $\hbar$ and $c$. We set $\hbar=c=1$ throughout the paper.}. Although we do not pursue this in detail in this paper, in Appendix B, we comment on how such rapid nonlinear fragmentation of the field, and to a lesser extent, the time-dependent evolution of oscillon configurations themselves could have interesting integrated Sachs-Wolfe (ISW) signatures. 

While similar resonant phenomenon are commonly invoked in the early Universe, for example, during preheating ($z\gg 10^{20}$) \cite{Allahverdi:2010xz}, few observational signatures exist (e.g. \cite{Khlebnikov:1997di,Easther:2006gt,Dufaux:2007pt,Khlopov:2008qy}). The reason for this lack of observational signatures on astrophysical scales is that resonance is effective only on subhorizon scales. Since high redshifts imply a small Hubble horizon, only small astrophysically inaccessible scales today (of order meters for grand uniÞed theory scale inflation) become excited. Moreover, on these small scales, the thermal radiation dominated state of the Universe at the time of big bang nucleosynthesis smooths out inhomogeneities. Since in our scenario the transition in quintessence is happening today, the resonance now happens on astrophysically accessible scales. As a result these models can be strongly constrained. 

The observationally interesting phenomenology of quintessence oscillations comes at a price. In addition to the usual problems with quintessence models (such as the smallness of the energy density, long range forces, etc.), for the quintessence potentials we work with, we require that the transition to an oscillatory regime happens close to today. This imposes an additional tuning for the initial conditions of the field. 

Related to the present paper \cite{Johnson:2008se}, Johnson and Kamionkowski discuss the dynamical instability arising from a ``small" anharmonic term in oscillating dark energy models and conclude that such instabilities render oscillating dark energy models unsuitable for providing a sustained cosmic acceleration period. While we rely on these very instabilities to source the rapid growth of the field fluctuations and the gravitational potential, we envision the transition happening late enough so that this instability does not significantly affect the observable expansion history. We then calculate and discuss in detail the additional observables that can be used to probe the rapid growth in the gravitational potentials arising from the dynamical instability. We reiterate, that we are not interested in the oscillatory phase providing the accelerated expansion. 

We briefly mention a few papers that explore clustering in quintessence-like fields, though through very different mechanisms. Observational effects of quintessence clustering caused by phenomenologically varying its speed of sound ({\it{not}} from resonant behavior discussed here) has been investigated by several authors (see, for example, \cite{DeDeo:2003te, Weller:2003hw, Bean:2003fb, Hu:2004yd,Rapetti:2004aa, Alimi:2009zk,dePutter:2010vy}). Instabilities in coupled dark energy-dark matter models have also been considered in \cite{Bean:2007ny}. In addition, rapid transitions in dark energy, in the context of coupled/unified dark matter and dark energy models where dark energy ``switches on"  at late times have been explored by \cite{Bertacca:2010mt} (also see references therein). In our case, dark energy ``turns off" at late times as the quintessence field starts oscillating, leading to a rapid growth in field fluctuations. With a somewhat different motivation, in \cite{Mortonson:2009qq} the authors discuss how rapid, extremely low redshift ($z\lesssim 0.02$) transitions in dark energy can be hidden from expansion history measurements. However, they did not consider the effects of such transitions on perturbations, which is the focus of this work. 

In this exploratory paper, we restrict ourselves to single field quintessence models, assume a minimal coupling to gravity and impose no nongravitational couplings to other fields. We include Weakly Interacting Massive Particle (WIMP) dark matter along with the quintessence field and assume a spatially flat universe. We limit ourselves to a linear treatment of the fluctuations, including scalar gravitational perturbations, except when we discuss the nonlinear fragmentation of the quintessence field and the formation of robust, localized quintessence excitations (oscillons). We focus on a case that is consistent with expansion history and galaxy clustering but could potentially be ruled out by measurements of the gravitational potential, in particular, via the cosmic microwave background (CMB) temperature power spectrum at large angular scales. Our intention is to stress the interesting phenomenology of such models and to point out that such transitions can be better constrained by going beyond the measurements of expansion history and growth of structure in galaxies alone. 

The rest of the paper is organized as follows. In Sec. \eqref{s:model} we introduce and motivate the form of the quintessence potential used in this paper. In Sec. \eqref{s:Hom} we work through the evolution of the quintessence field in an Friedmann-Robertson-Walker universe. We highlight initial conditions and important regimes of evolution as well as constraints placed on the parameters from observations of the expansion history. In Sec.  \eqref{s:Fluc} we discuss the evolution of linearized fluctuations in the quintessence field, dark matter and the gravitational potential, with a special emphasis on their evolution during the oscillatory phase of the background quintessence field. In the same section we also discuss the domain of validity of our linearized treatment and comment on the nonlinear evolution of the field. In Sec. \eqref{s:Obs} we compute observables such as lensing power spectra and the ISW contribution to the CMB temperature anisotropy. We present our conclusions in Sec. \eqref{s:Con}. We also include two Appendixes. In Appendix A, we provide some details of Floquet analysis and an algorithm used in this paper for calculating Floquet exponents. In Appendix B, we provide an estimate of the ISW effect resulting from evolution of quasi/nonlinear quintessence fluctuations and oscillons.

\section{The model}
\label{s:model}

\begin{figure}[t] 
   \centering
   \includegraphics[width=4in]{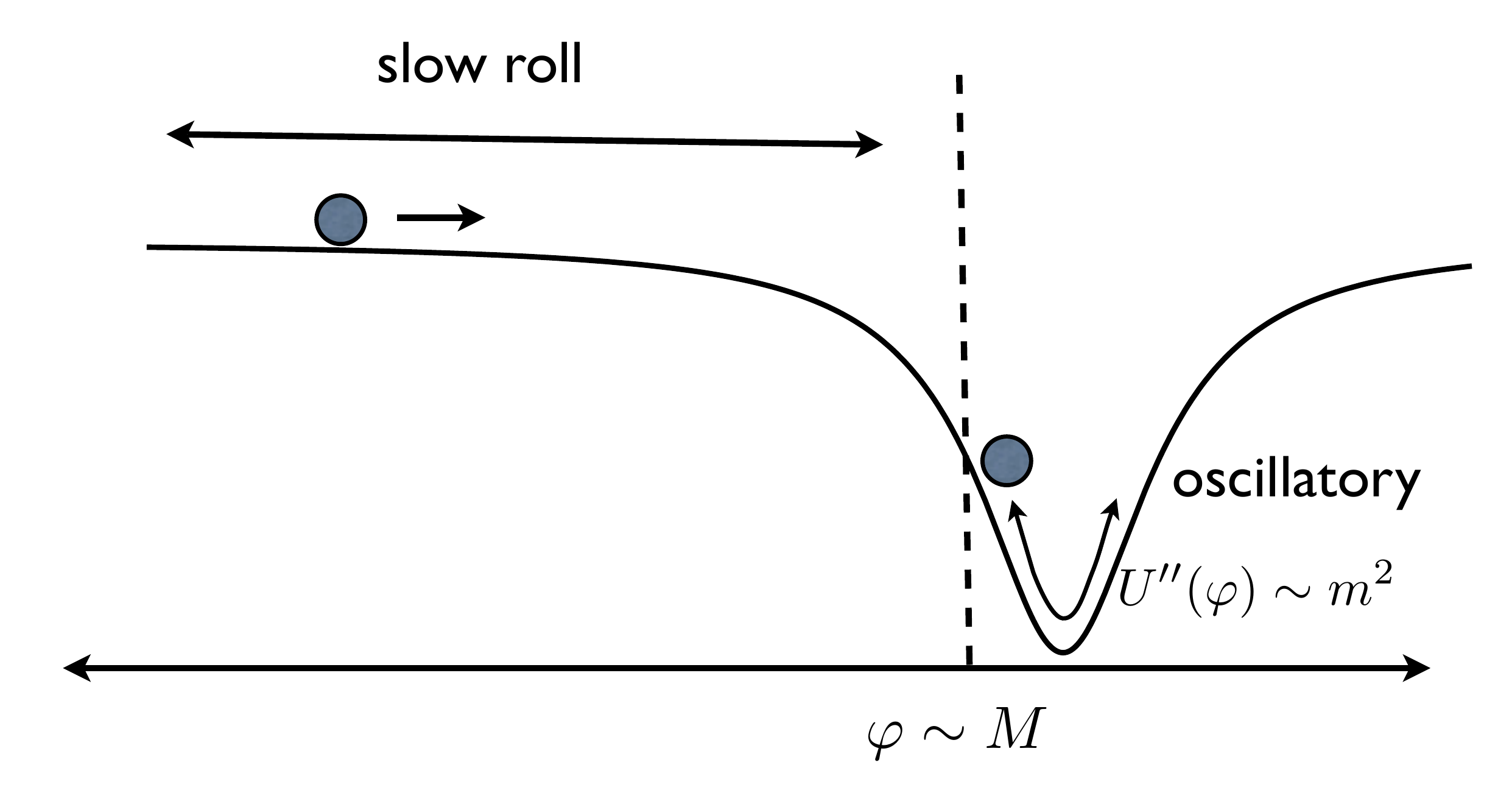} 
   \caption{Initially, the field rolls slowly  causing accelerated expansion of space. As it enters the oscillatory regime, the acceleration stops and a rapid, exponential growth of field fluctuations begins.}
   \label{fig:potential}
\end{figure}

\noindent We consider a quintessence field governed by a potential of the form (see Fig. \ref{fig:potential}):
\Beq
\label{eq:potentialAlpha}
U(\vp)=\frac{m^2M^2}{2}\left[\frac{(\vp/M)^2}{1+(\vp/M)^{2(1-\alpha})}\right],
\Eeq
where $0<\alpha<1$. \footnote{We write $\vp$ instead of $|\vp|$ to avoid clutter.}  This choice was motivated by monodromy and supergravity models of inflation \cite{Silverstein:2008sg,McAllister:2008hb, Flauger:2009ab,Kallosh:2010ug,Dong:2010in} and a recent model of axion quintessence \cite{Panda:2010uq} ($\alpha=1/2$). The potential has a quadratic minimum and for very large field values it asymptotes to a shallower than quadratic form,
\Beq
&U(\vp)\approx \frac{m^2}{2}\vp^2\quad\quad &\vp\ll M,\\
&U(\vp)\approx \frac{m^2M^2}{2}(\vp/M)^{2\alpha}\quad\quad &\vp \gg M.\\
\Eeq
The scale $M$ determines where the potential changes shape, whereas, the scale $m$ determines the curvature $U''(0)$ at the bottom of the well. We will consider cases where the field rolls slowly for $\vp\gg M$, behaving like dark energy and then  enters an oscillatory regime with $\vp\sim M$ after $z\sim 0.2$. When the field oscillates around the minimum, the ``opening up" of the potential,
\Beq
\label{eq:condition}
&U(\vp)-(1/2)m^2\vp^2<0\quad\quad \vp \ne 0,
\Eeq
leads to rapid, scale dependent growth of scalar field fluctuations  via parametric resonance (see Sec. \eqref{s:Fluc}). In the nonlinear regimes, it leads to the formation of localized field excitations called oscillons \cite{Amin:2010xe,Amin:2010jq}.

Similar phenomenology can arise in models such as pseudo-Nambu-Goldston-Boson quintessence (e.g. \cite{Hall:2005xb}) 
\Beq
\label{eq:potentialAxion}
U(\vp)=m^2M^2\left[1-\cos(\vp/M)\right].
\Eeq
For potentials \eqref{eq:potentialAlpha} and \eqref{eq:potentialAxion}, a limited range of parameters will reproduce the observed expansion history, allow for a few oscillations in the field close to today, {\it and} allow for rapid growth of structure. For the pseudo-Nambu-Goldston-Boson-like models, the slow roll dynamics of the field necessary for accelerated expansion, only occur if $M\sim \mpl$ (unless $\vp \rightarrow \pi M$). However, as we discuss in Sec. \eqref{s:Fluc}, efficient resonance in an expanding universe requires $M\ll \mpl$.  As a result we do not get rapid growth of structure here, except in cases of extreme fine tuning of the initial conditions and do not pursue this model any further in this paper.

For the potential in \eqref{eq:potentialAlpha}, we find the requirement of a few oscillations close to today translates into
\Beq
10^2H_0\lesssim M,m\lesssim 10^{-2}\mpl \quad\textrm{and}\quad \alpha\ll 1.
\Eeq 
We study these constraints in more detail in the next section.  
 
Finally, we note that resonant growth of fluctuations also arise when scalar fields oscillate in potentials with asymmetric minima
$$U(\vp)\approx \frac{m^2}{2}{\vp^2}+\frac{M}{3}\vp^3+\hdots,$$ and is possible in potentials with a nonquadratic minimum or in potentials, which do not necessarily open up (though the band of resonant wave-numbers can be narrow). The key requirement is that there are anharmonic terms in the potential that provide a time-varying, periodic frequency in the equation of motion for the fluctuations. 


\section{Homogeneous evolution}
\label{s:Hom}

In this section, we describe how to choose initial conditions for the background quintessence field and parameters in the potential so that the field both oscillates at late times and reproduces the observed expansion history. We do this by trying to match the $\Lambda$CDM  expansion history. This is more restrictive than directly using observations but makes the following discussion more transparent. 


The equations of motion for the homogeneous quintessence field are 

\Beq
\label{eq:Hom}
&\ddot{\vp}+3H\dot{\vp}+U'(\vp)=0,\\
&H^2=\frac{1}{3\mpl^2}\left[\frac{\dot{\vp}^2}{2}+U(\vp)\right]+H^2_0\frac{\Omega_{\rm{dm}}}{a^3},
\Eeq
where $\Omega_{\rm{dm}}$ is the fraction of the critical energy density in WIMPs today, $H$ is the Hubble parameter and $H_0$ is its value today \footnote{We choose the subscript ``dm" for WIMPs instead of the more usual $m$ to avoid confusion with the mass of the scalar field} .  The dot ``.'' stands for a derivative with respect to cosmic time and the scale factor $a=1$ today. We ignore the contribution from radiation since we are only interested in late times.  

Current observations are consistent with the quintessence field behaving like a cosmological constant in the recent past \cite{Serra:2009yp}. During the matter dominated epoch, it is more difficult to place constraints on dark energy's equation of state because it is subdominant. For simplicity, we assume that the field's equation of state satisfies $w\equiv (\dot{\vp}^2-2U)/(\dot{\vp}^2+2U)\sim-1$ (for $z\gtrsim 0.2$), which occurs if the `slow roll' condition $\dot{\vp}^2\ll U$ is satisfied. Imposing that the energy density in the quintessence field during matter domination has the same order of magnitude as the $\rho_{\Lambda}$ in $\Lambda$CDM, we find,

\Beq
\label{eq:constraint1}
\left(\frac{m}{H_0}\right)^{\!2}\left(\frac{\vp_i}{M}\right)^{\!2\alpha}\sim6(1-\Omega_{\rm{dm}})\left(\frac{\mpl}{M}\right)^{\!2}
\Eeq

\noindent where $\vp_i$ is the initial value of the field deep in the matter dominated epoch. Note that for simplicity, we have assumed that $\vp_i\gg M$ and later justify our assumption at the end of the analysis. Since we are only interested in scaling relationships, we will ignore factors of order unity from now on. The above gives one constraint between parameters in the potential and the field's initial conditions. 

Imposing oscillations in the field at late times gives a second constraint. To derive this, we solve Eq.\;\eqref{eq:Hom} in the slow roll, matter dominated regime, assuming the field does not change significantly from its initial value. We find (assuming $\vp_i\gg M$),\footnote{The apparent discontinuity between the cases $\alpha=0$ and $\alpha>0$ disappears when the full expressions are used.}

\Beq
\label{eq:vpSol}
\frac{\vp}{M}-\frac{\vp_i}{M}\sim
\begin{cases} 
-a^3(m/H_0)^2(M/\vp_i)^{3}& \text{if $\alpha=0$}, \\ 
-a^3(m/H_0)^2(M/\vp_i)^{1-2\alpha}\alpha& \text{if $\alpha>0$}.
\end{cases}
\Eeq

\noindent Though the above equation of motion is not valid in the oscillation regime, it gives an approximate scale factor $(a_*)$ when oscillations begin ($\vp/M\sim 0$). We find,

\Beq
\label{eq:constraint2}
a_*\sim
\begin{cases}
(H_0/m)^{2/3}(\vp_i/M)^\frac{4}{3}& \text{if $\alpha=0$}, \\ 
(H_0/m)^{2/3}(\vp_i/M)^\frac{2-2\alpha}{3}\alpha^{-\frac{1}{3}}& \text{if $\alpha>0$}.\\
\end{cases}
\Eeq

\noindent The actual scale factor when oscillations begin ($a_{\rm{osc}}$) differs from $a_*$ but the two are monotonically related. Combining Eqs.\;\ref{eq:constraint1} and \;\ref{eq:constraint2}, we calculate explicit expressions for $m$ and $\vp_i$ in terms of $M$ and $a_{*}$. We have,

\Beq
\label{eq:constraints}
\frac{\vp_i}{M}&\sim
\begin{cases} 
a^{\frac{3}{4}}_*(\mpl/M)^{\frac{1}{2}} & \text{if $\alpha=0$}, \\ 
a^{\frac{3}{2}}_*(\mpl/M)\alpha^{\frac{1}{2}} & \text{if $\alpha>0$}. 
\end{cases} \\
\frac{m}{H_0}&\sim
\begin{cases} 
(\mpl/M) & \text{if $\alpha=0$}, \\ 
a^{\frac{-3\alpha}{2}}_*(\mpl/M)^{1-\alpha}\alpha^{-\frac{\alpha}{2}}  & \text{if $\alpha>0$}. 
\end{cases}
\Eeq

As discussed in more detail below (in Sec. \ref{s:Fluc}\;), $M$ sets the strength of the resonance between the perturbed quintessence field and the background quintessence field. We therefore keep it as a free parameter. We choose $a_*$ so that the actual numerically computed oscillations begin around $a_{\rm{osc}}\approx 0.8$. Corrections to distance measurements that depend on cosmological parameters beyond $H_0$ are small for scale factors larger than $a_{\rm{osc}}\approx 0.8$. Hence, modifying the expansion history there will not interfere with measurements that probe the cosmological expansion. Finally, we fix $m$ and $\vp_i$ using Eq.\;\eqref{eq:constraints}. The value of $\dot{\vp}_i$ follows from Eq.\;\eqref{eq:vpSol}. 

\begin{figure}[H] 
   \centering
   \includegraphics[width=5in]{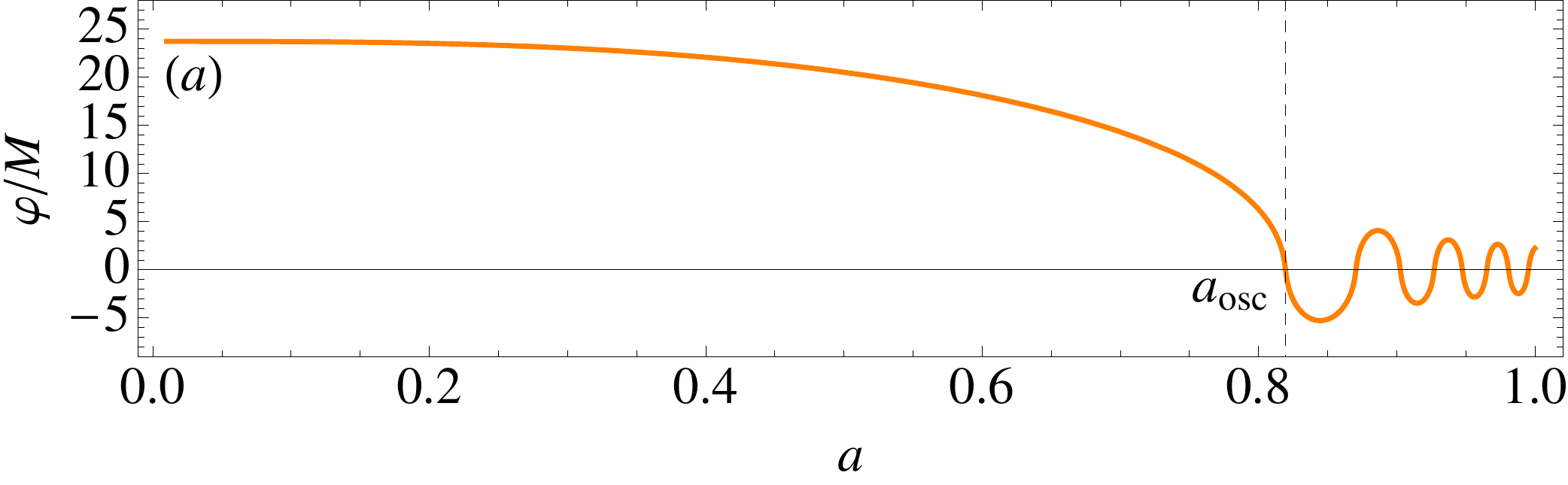} \\
    \includegraphics[width=5in]{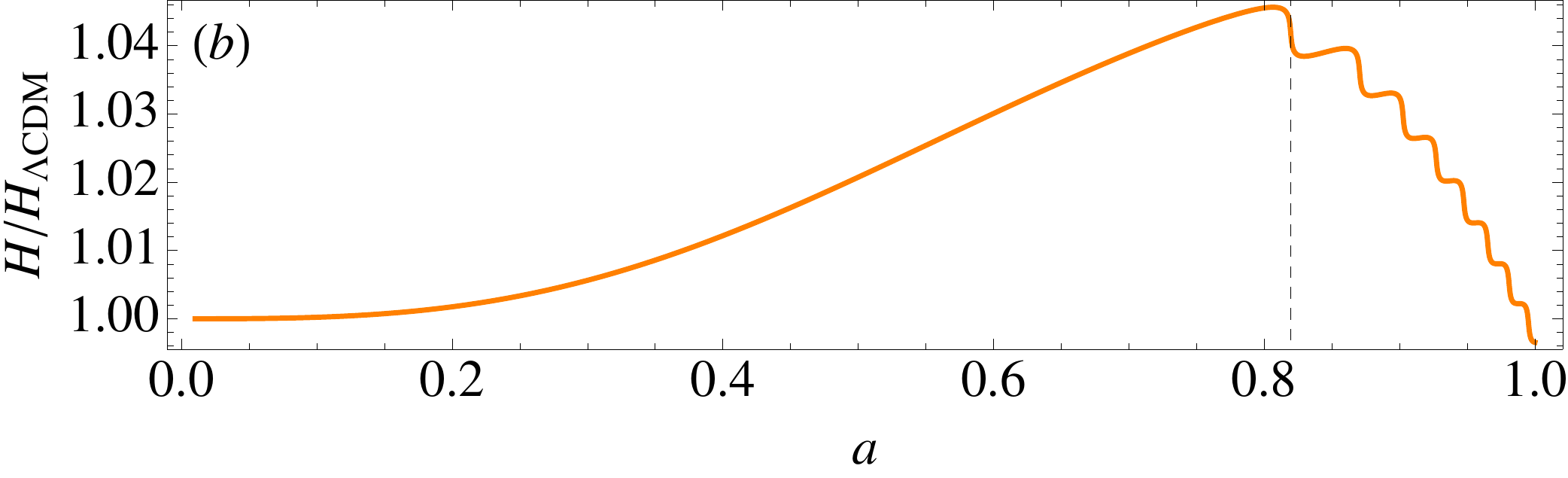} \\
    \includegraphics[width=5in]{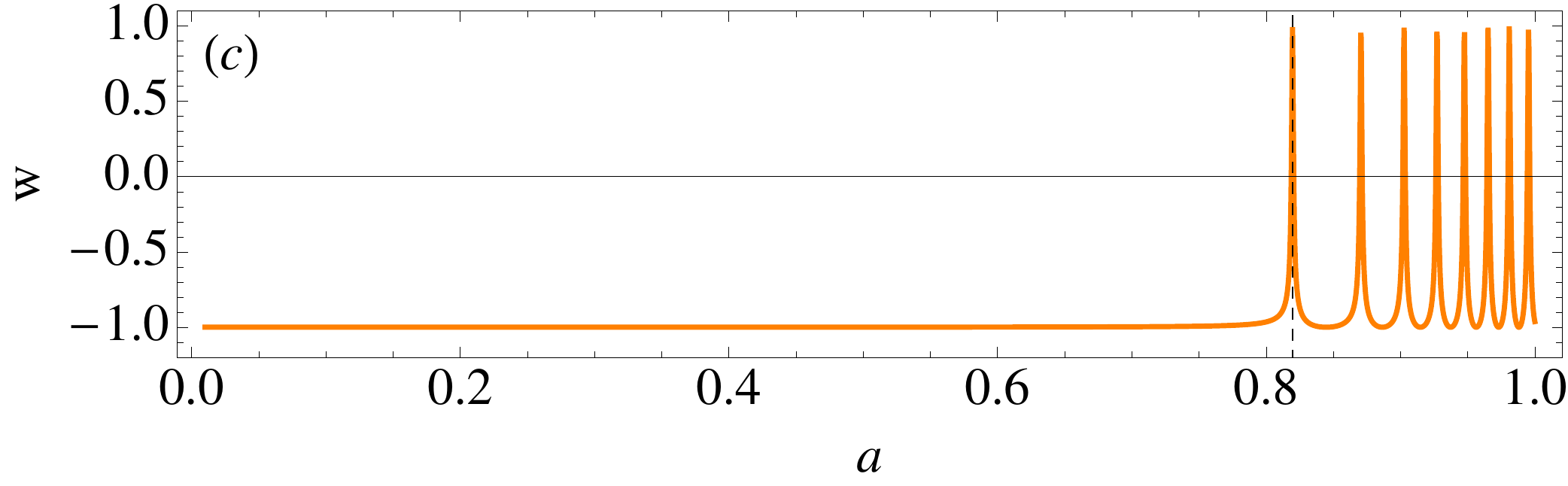} \\
        \includegraphics[width=5in]{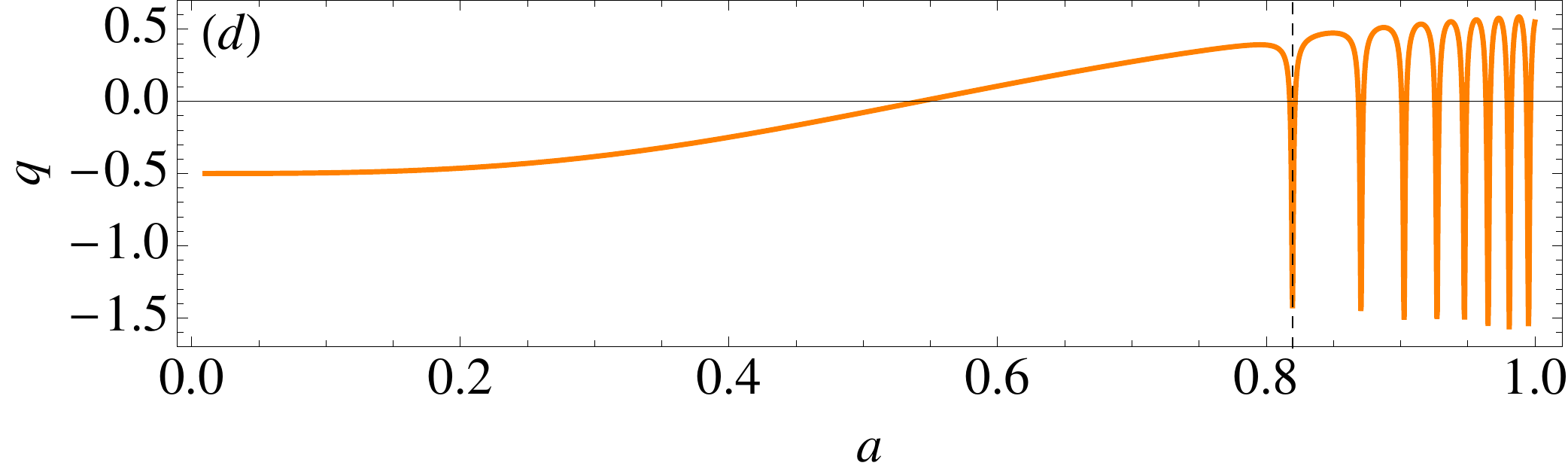}

   \caption{(a) Homogeneous field evolution: note that the field is relatively constant until $a_{osc}$ after which it starts oscillating with a decaying amplitude. (b) The Hubble parameter increases initially compared to its $\Lambda$CDM counterpart but once the oscillations in the field begin, it starts decreasing again. The small oscillations in the Hubble parameter reflect the oscillations in the quintessence field. (c) The equation of state parameter $w\approx-1$ until $a_{\rm{osc}}$, reflecting the cosmological constant like behavior due to the slowly rolling field, but starts oscillating about $w=0$ after $a_{\rm{osc}}$, reflecting the approximately nonrelativistic matter like behavior of the oscillating field. (d) The acceleration parameter $q$  reveals a period of decelerated expansion in the matter dominated era, followed by a period of accelerated expansion as the slowly rolling quintessence field begins to dominate the energy density. This continues until $a_{\rm{osc}}$, after which $q$ starts oscillating due to the oscillations in the quintessence field. }
   \label{fig:Hom}
\end{figure}

In practice, when numerically evolving Eq.\;\eqref{eq:Hom} according to the above choices of initial conditions and potential parameters, we find that in order to get agreement with the \lcdm expansion history in the past as well as produce oscillations close to today, we require additional fine tuning of the initial conditions and parameters. The procedure described above results in a value of $H_0$ that is smaller than the \lcdm value. This is because by assumption, the Hubble parameter for both \lcdm and quintessence are identical. At late times, however, the oscillations in the quintessence field cause it to behave like dark matter, giving rise to a steeper fall off than the \lcdm Hubble parameter. For better agreement with $H_0$, we increase the value of $m$ until $H_0$ agrees to better than a percent. 

Performing the above iteration, we find that $\alpha$ affects whether or not the field starts oscillating at low redshift. This should be expected since the asymptotic slope of the potential determines the time it takes to transition from slow-roll to oscillations. We can calculate this approximate time from the derivative of the quintessence field when slow roll ends. Using the slow roll condition described above, the value of the quintessence field when slow roll ends is, 

\Beq
\vp_e\sim
\begin{cases}
(\mpl/M)^{1/3}M& \text{if $\alpha=0$}, \\ 
\alpha\mpl  & \text{if $\alpha>0$}. 
\end{cases}
\Eeq

\noindent Assuming $3H\dot{\vp}\sim-U'$, taking this transition to happen close to today, and using the constraints in Eq.\;\eqref{eq:constraints}, we find the change in scale factor ($\triangle a$) between the beginning of oscillations ($\vp\sim0$) and the end of slow roll is given by,

\Beq
\label{eq:alpha}
\triangle a\sim
\begin{cases}
\left(M/\mpl\right)^{2/3}& \text{if $\alpha=0$}, \\ 
\alpha^{1-\alpha}  & \text{if $\alpha>0$}. 
\end{cases}
\Eeq

\noindent The larger the $\alpha$, the longer it takes for oscillations to start. Hence, consistency with a \lcdm expansion history and having late-time oscillations limits $\alpha\ll1$ for the case of nonzero $\alpha$ and $M\ll\mpl$ for vanishing $\alpha$. For the rest of this paper, we specialize to $\alpha=0$. In summary, we keep $M$ and $a_{*}$ as free parameters and choose $m$, $\vp_i$ and $\dot{\vp}_i$ so that the value of $H_0$ agrees with the measured value, the field behaves like dark energy initially, and the field starts oscillating at $a_{\rm{osc}}\approx 0.8$.

For the above prescription, we calculate the evolution of the quintessence field, the Hubble parameter normalized by $\Lambda$CDM's Hubble parameter, the field's equation of state parameter $w$ and the acceleration parameter $q\equiv\ddot{a}a/\dot{a}^2$ as a function of scale factor. We show the results in Fig. \ref{fig:Hom} for 
\Beq
\mpl/M=500,\,m/H_0=1130.6,\,\vp_i/M=23.7\,\,{\rm{with}}\,\,\,a_{\rm{osc}}=0.82.
\Eeq Unless otherwise stated we will use these as fiducial parameter values throughout this paper and refer to them as fiducial parameters. As constructed, the field remains approximately constant for $a\ll a_{\rm{osc}}$, rolling slowly towards the minimum. Notice, as assumed initially, that $\vp\gg M$ at early times. At $a_{\rm{osc}}\sim 0.8$, the field starts to oscillate with a slowly decaying amplitude,

\Beq
\vp\approx\vp_{\rm{osc}}(a)\sin(\omega t+\Delta).
\Eeq

\noindent Note that when $\vp\ll M$, $\vp_{\rm{osc}}(a)\propto a^{-3/2}$. Because of the anharmonic terms in the potential, the frequency of oscillation $(\omega)$ depends on the amplitude of the field and is given by,

\Beq
\label{T}
\omega\approx m\left\{\frac{2}{\pi}\frac{1}{\sqrt{1+(\vp_{\rm{osc}}/M)^2}\;\;{\cal{E}}\left[-(\vp_{\rm{osc}}/M)^2\right]}\right\}\le m,
\Eeq

\noindent where ${\cal{E}}$ is the complete elliptic integral. As $\vp\rightarrow0$, $\omega\rightarrow m$. Note that in deriving the above expression for the period, we have assumed $\omega\gg H_0$ so that energy is approximately conserved during an oscillation.  
 
The Hubble parameter matches \lcdm at early times since both models have the same amount of dark matter [see Fig. \ref{fig:Hom}(b)]. The Hubble parameter must then increase relative to \lcdm in order for $H_0$ to agree, since the quintessence model's Hubble parameter falls off more steeply than \lcdm at late times when the field starts to behave like nonrelativistic matter. The deviation, for the fiducial parameters, is at most 5\%, which is consistent with recent observational constraints \cite{Blake:2011ep}. The equation of state parameter behaves  as expected in the slow roll regime ($w=-1$) and oscillates between 1 and -1 after $a_{\rm{osc}}$ as energy swaps between its kinetic and potential parts [Fig. \ref{fig:Hom}(c)]. Note that other models that give rise to an oscillating equation of state have been explored in the literature (for example, \cite{Linder:2005dw,Dutta:2008px,Lazkoz:2010gz,Dutta:2006pn,Amendola:2007yx,Tarrant:2011qe,Baldi:2011th}). Lastly, the acceleration parameter reveals a period of decelerated expansion deep in the matter dominated era, followed by a period of accelerated expansion as the slowly rolling quintessence field begins to dominate the energy density.  After $a_{\rm{osc}}$, the acceleration parameter starts oscillating around $q=-1/2$ [Fig. \ref{fig:Hom}(d)]. 
\begin{figure}[t] 
   \centering
   \includegraphics[width=5in]{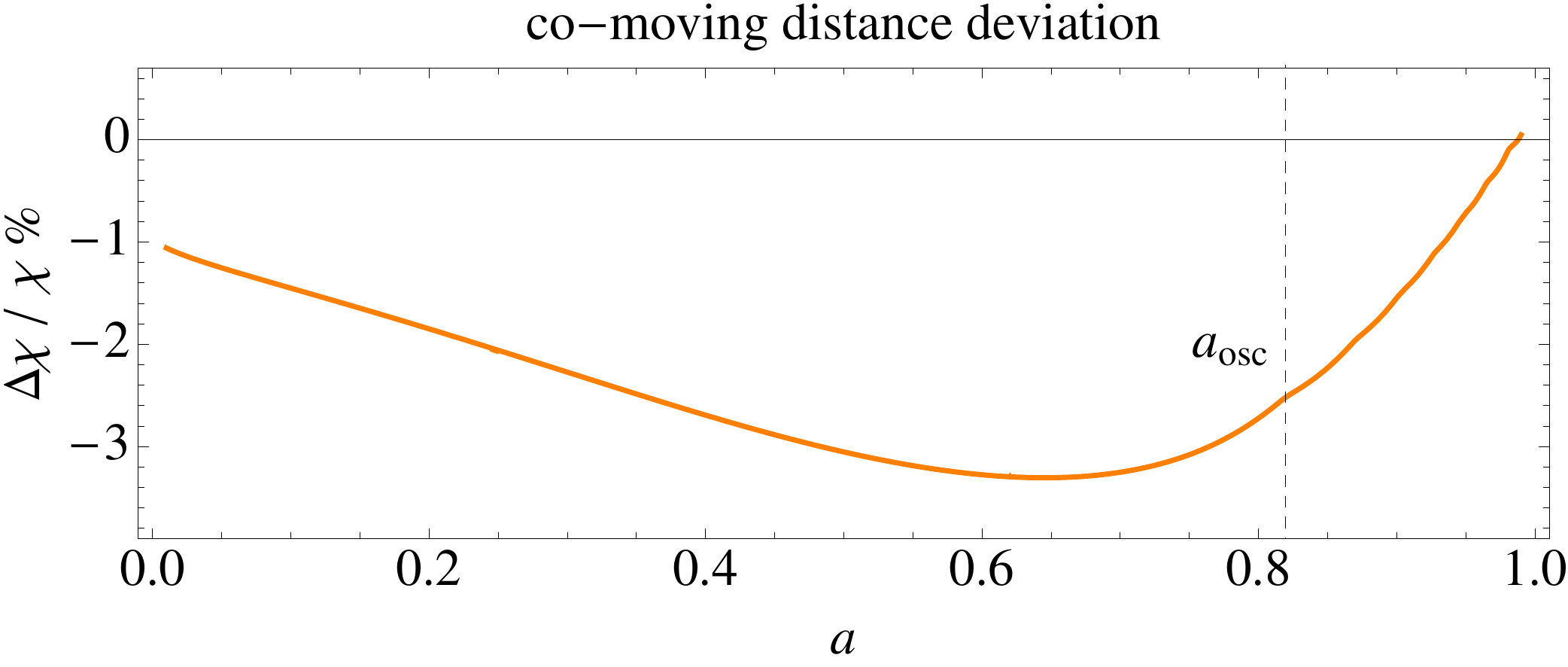} \\
   \caption{The difference in comoving distance from an observer today, between the quintessence and \lcdm models. Deviations are smaller today, where both cosmologies have the same $H_0$, and at early times, where both cosmologies have the same amount of dark matter and dark energy.}
   \label{fig:ChiDiff}
\end{figure}

Observational constraints on the expansion history at late times arise primarily from distance measurements. Given our above prescription, we expect good agreement between distances calculated in our quintessence model and $\Lambda$CDM, which we show explicitly below. The comoving distance is given by,

\Beq
\chi(a)=\int^1_a\frac{da}{a^2H(a)}.
\Eeq

In Fig. \ref{fig:ChiDiff}, we plot the percent difference in $\chi(a)$ between quintessence and $\Lambda$CDM. As expected, deviations are smaller close to today, where the Hubble parameter was tuned to be identical to the observed $H_0$ and at early times, where dark energy can be ignored. The largest deviations are $\sim4\%$. Since all observed distances, like the luminosity distance and the angular diameter distance, differ from the comoving distance by factors of the scale factor, Fig. \ref{fig:ChiDiff} implies that current observations cannot distinguish between the background evolution of quintessence and the background evolution in $\Lambda$CDM \cite{Komatsu:2010fb,Rapetti:2006fv,Davis:2007na,Riess:2009pu,Kessler:2009ys}. We have restricted our analysis to $a_{\rm{osc}}\sim 0.8$ because choosing $a_{\rm{osc}}\ll 0.8$ gives rise to unacceptably large deviations  from the observed expansion history. 


\section{Perturbation evolution}
\label{s:Fluc}
In this section we investigate the dynamics of the fluctuations in the quintessence field ($\delta\vp$), the gravitational potentials, and the overdensity in WIMPs ($\delta_{\textrm{dm}}$) assuming linearized equations of motion. We ignore radiation and neutrinos, since we are interested in late-time dynamics. We work in the Newtonian gauge, where the metric is
\Beq
ds^2=-(1+2\Phi)dt^2+a^2(1-2\Psi)dx^2,
\Eeq
and $\Phi$ and $\Psi$ are the two scalar gravitational potentials. Tracking the evolution of the gravitational potentials and the two matter components (WIMPs and quintessence) requires two second order differential equations. Normally, they are taken to be the energy-momentum conservation equations for the matter components. The gravitational potentials can then be obtained from the Einstein equations (constraints). However, since the most interesting dynamics happen in the gravitational potentials and the field fluctuations, we work with them as the degrees of freedom and then obtain the overdensity in WIMPs from the constraints. The equations of motion for the quintessence field fluctuations and the gravitational potential are (in Fourier space),
\Beq
\label{eq:PertEOM}
&\ddot{\delta\vp}_k+3H\dot{\delta\vp}_k+\left[\frac{k^2}{a^2}+U''(\vp)\right]\delta\vp_k=-2U'(\vp)\Psi_k+4\dot{\vp}\dot{\Psi}_k,\\
&\ddot{\Psi}_k+4H\dot{\Psi}_k+\frac{1}{\mpl^2}U(\vp)\Psi_k=\frac{1}{2\mpl^2}\left[\dot{\vp}\dot{\delta\vp}_k-U'(\vp)\delta\vp_k\right].
\Eeq
The second equation is the diagonal, space-space component of the Einstein equations, where the right hand side is the pressure perturbation provided by the scalar field. Since the WIMP dark matter is assumed to be pressureless, it does not contribute to this source term.  The second gravitational potential $\Phi_k$ does not appear in the above equations because  there is no anisotropic stress at linear order for single, minimally coupled scalar fields. This makes the two gravitational potentials equal 

\begin{equation}
\Psi_k=\Phi_k,
\end{equation}

\noindent via the off diagonal space-space part of the Einstein equations.{\footnote{For nonminimally coupled scalar fields, $\Phi_k-\Psi_k\propto f'(\vp)/f(\vp)\delta\varphi_k$ where $8\pi G_N\rightarrow f(\varphi)^{-1}$. In minimally coupled fields, the anisotropic stress is generated by second order terms: $\Phi_k-\Psi_k\propto \delta \vp_k^2$.} 
After evolving the equations for $\delta\vp_k$ and $\Psi_k$, $\delta_{\textrm{dm}}$ follows from the time-time components of the Einstein equations,
 \Beq
 \label{eq:dmConstraint}
 \delta_{\textrm{dm}}=-\frac{a^3}{3H_0^2\Omega_{\textrm{dm}}}\left[\left(6H^2-\frac{\dot{\vp}^2}{\mpl^2}+2\frac{k^2}{a^2}\right)\Psi_k+6H\dot{\Psi}_k+\frac{\dot{\vp}}{\mpl^2}\dot{\delta\vp}_k+\frac{1}{\mpl^2}U'(\vp)\delta\vp_k\right].
 \Eeq

 \subsection{Initial conditions and evolution during matter domination }
 \label{ss:MDPert}
 We are interested in the behavior of $\delta\vp_k$, $\Psi_k$ and $\delta_{\textrm{dm}}$ for $a\gtrsim 0.8$. To solve \eqref{eq:PertEOM} we need to specify $\delta\vp_k,\delta\dot{\vp}_k,\Psi_k,\dot{\Psi}_k$ on some initial time slice. This can be done self-consistently using the $\Lambda$CDM solutions for $\Psi_k$ and $\delta_{\rm{dm}}$ if the initial time slice is chosen deep into the matter dominated era. During matter domination, the quintessence field is a small fraction of the total energy density. As a result, the $\delta\vp_k$ do not contribute significantly to $\Psi_k$, and we can use $\Psi_k$  from a $\Lambda$CDM cosmology at these early times. We take these initial conditions for $\Psi_k$ and $\dot{\Psi}_k$ directly from the output of CMBFAST \cite{Seljak:1996is} at $a_i\approx 10^{-2}$ since by this scalefactor the anisotrotropic stress from neutrinos is negligible and the contribution of dark energy is yet to become important. We are then left with specifying the initial conditions in $\delta\vp_k$, which requires understanding its evolution in the matter dominated era.
 
The evolution of adiabatic modes on superhorizon $k/aH\ll 1$ scales is given by (see \cite{Weinberg:2008zzc})\footnote{We will ignore isocurvature modes in this paper}
\Beq
\label{eq:SH}
\delta\varphi_k=\Psi_k\frac{\dot{\vp}}{H}\left[\frac{(H/a)\int_{t_i}^t a(t')dt'}{-1+(H/a)\int_{t_i}^t a(t')dt'}\right].
\Eeq
The term in the square brackets is constant during matter domination
\Beq
\label{eq:SHmatter}
& \delta\varphi_k=-\frac{2}{3}\Psi_k\frac{\dot{\vp}}{H}.
 \Eeq

On subhorizon scales ($k\gg aH$), during the matter dominated era, the gravitational potential is determined by the fluctuations in the WIMP overdensity. 
Since $\Psi_k$ is constant during matter domination, the equations of motion \eqref{eq:PertEOM} become
 \Beq
&\ddot{\delta\vp}_k+3H\dot{\delta\vp}_k+\frac{k^2}{a^2}\delta\vp_k=-2U'(\vp)\Psi_k,\\
&\dot{\Psi}_k=0,
\Eeq
where we have assumed $U''(\vp)/H^2\ll (k/aH)^2$. Under the assumption that $U'(\vp)$ is slowly varying,  the complete solution is

\Beq
\label{eq:sHmatterF}
\delta\vp_k=
& \frac{c_k}{k_H^{3}}\Big[\cos (2k_H+\theta_k)+2k_H\sin(2k_H+\theta_k)\Big]-2{\Psi_k}\frac{U'(\vp)}{H^2}\frac{1}{k_H^2}\left[1-\frac{7}{k_H^2}+\frac{35}{2k_H^4}\right],
\Eeq
where $k_H\equiv k/aH$ and $c_k, \theta_k$ are constants of integration set by initial conditions at the beginning of the matter dominated era. For solutions that are deep inside the horizon, $k_H\gg1$ and only one term grows with time:
\Beq
\label{eq:sHmatter}
\delta\vp_k\approx-{2\Psi_k\frac{U'(\vp)}{H^2}}\frac{1}{k_H^2}.
\Eeq
The initial (transient) oscillatory behavior due to the homogeneous part of the solution \ref{eq:sHmatterF} as well as the  growth $\propto a^2$ for $a_i\ll a\ll \aosc$ due to the particular solution \ref{eq:sHmatter} can be seen in Fig. \ref{fig:ResGrowth}(a). We have also verified that although from Eqs.\;\eqref{eq:SHmatter} and \eqref{eq:sHmatter} we see that $\delta\vp_k$ grows as $a^3$ on superhorizon scales and as $a^2$ on small, subhorizon scales, their amplitude during matter domination does not become large enough to change the behavior of the potential $\Psi_k$.


With the Eqs.\;\eqref{eq:SHmatter} and \eqref{eq:sHmatter} at hand, we set the initial conditions for all $k_H$ using a piecewise interpolation between the subhorizon and superhorizon solutions (dashed curve in Fig. \ref{fig:SpectraField}). Although we have taken care to faithfully characterize the asymptotic behavior of the initial conditions of $\delta\vp$, in practice we find that changing the initial conditions by orders of magnitude does not affect our results significantly. This is because at late times, the particular solution of $\delta\vp_k$ (determined by $\Psi_k$) dominates, significantly reducing the dependence on initial conditions.  


 \subsection{Evolution during the quintessence dominated era}
In this section, we investigate the dynamics of fluctuations during quintessence domination and find that the evolution differs before and after $\aosc$. We analyze these two regimes separately. 
 
\subsubsection{Before resonance $a<\aosc$}
When $0.5<a<\aosc$, the homogeneous quintessence field rolls slowly with an energy density larger than, but comparable, to the WIMP density. The behavior of $\delta\vp_k$ on superhorizon scales is still determined by \eqref{eq:SH}, but $H$ decreases more slowly than during the matter dominated era. 

In the subhorizon regime, the quintessence perturbations grow faster as we approach $\aosc$ because the field starts rolling more rapidly (see right-hand side of Eq. \eqref{eq:PertEOM}). However, these fluctuations are still not strong enough to prevent the decay of the gravitational potential, which is caused by a transition to a dark energy dominated epoch. During this epoch, the evolution of the $\Psi_k$ is determined by $H$ and does not show any resonant behavior. The behavior of the $\delta\vp_k$ and $\Psi_k$ discussed here can be easily seen in Fig. \ref{fig:ResGrowth}(a) and (b) for $0.5\lesssim a\lesssim \aosc$. The thin, black line represents the evolution of the same mode of $\Psi_k$ in $\Lambda$CDM. 
 \subsubsection{Resonance $a>\aosc$}
We now come to the most interesting era ($a>\aosc$) with regards to the evolution of fluctuations. In this era, the homogeneous field begins to oscillate with an amplitude dependent frequency $\omega<m$ (see Eq. \eqref{T}). This leads to a rapid growth in the field fluctuations for certain characteristic ranges of wavenumbers. To understand this, we initially ignore expansion ($H=0, a=1$) and the gravitational perturbations ($\Psi_k=0$) in Eq. \eqref{eq:PertEOM}. 

\subsubsection*{Resonance in Minkowski space}
The equation of motion for $\delta\vp_k$ then becomes, 
\Beq
\label{eq:flatFloq}
\ddot{\delta\vp}_k+\left[k^2+U''(\vp)\right]\delta\vp_k=0.
\Eeq
Since the homogeneous field $\vp$ is periodic in time, $U''(\vp)$ is periodic as well as long as  $U''(\vp)\ne \rm{constant}$, which occurs for potentials with anharmonic terms (as is the case with our potential in \eqref{eq:potentialAlpha}). This  yields an oscillator with a periodically varying frequency, whose solutions can be analyzed via standard Floquet methods (for example, see \cite{Hill:1979}). For the interested reader, we review the main aspects of Floquet analysis in the Appendix. Under certain conditions (see Appendix A), Floquet's theorem guarantees that the general solution to Eq. \eqref{eq:flatFloq} can be written as 
\begin{equation}
\delta\varphi_k(t)= e^{\mu_k t}P_{+}(t)+e^{-\mu_k t}P_{-}(t),
\end{equation}
where $\pm\mu_k$ are the Floquet exponents and $P_{\pm}(t)$ are periodic functions with the same period as $U''(\vp)$.\footnote{When the Floquet exponents are $zero$, there exist another class of solutions $\delta\varphi_k(t)\propto t P_1(t)$ and $\delta\varphi_k(t)\propto P_2(t)$ where $P_{1,2}(t)$ are periodic functions.} The Floquet exponent $\mu_k$ depends on the amplitude of the ``pump" field $\vp_{\rm{osc}}$, as well as the wavenumber $k$. There exists an unstable, exponentially growing solution if the real part of the Floquet exponent $\Re(\mu_k)\ne0$.  In Fig. \ref{fig:Floquet}(a), we show  $|\Re(\mu_k)|$ as a  function of the amplitude $\vp_{\rm{osc}}$ and wavenumber. The color represents the magnitude of the real part of the Floquet exponent. Yellow corresponds to a larger $|\Re(\mu_k)|$ than orange, whereas red corresponds to $\Re(\mu_k)=0$. Without expansion, neither $\vp_{\rm{osc}}$ nor the momentum $k$ redshift with time. As a result, the evolution of modes is determined by the Floquet exponent at single point in the $(k,\vp_{\rm{osc}})$ plane. This no longer holds true in an expanding universe.
  \begin{figure}[t]
\begin{center}$
\begin{array}{cc}
 \includegraphics[width=3.in]{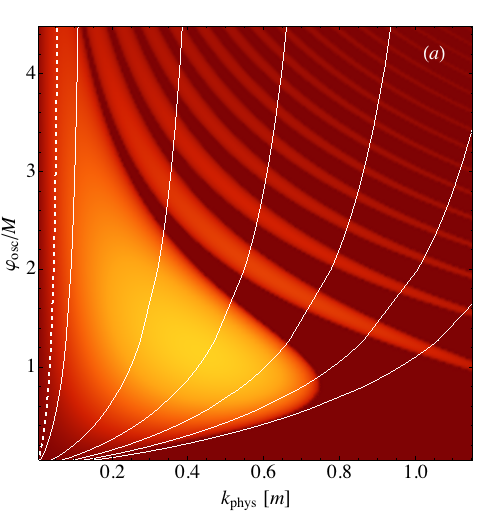} &
 \includegraphics[width=3.2in]{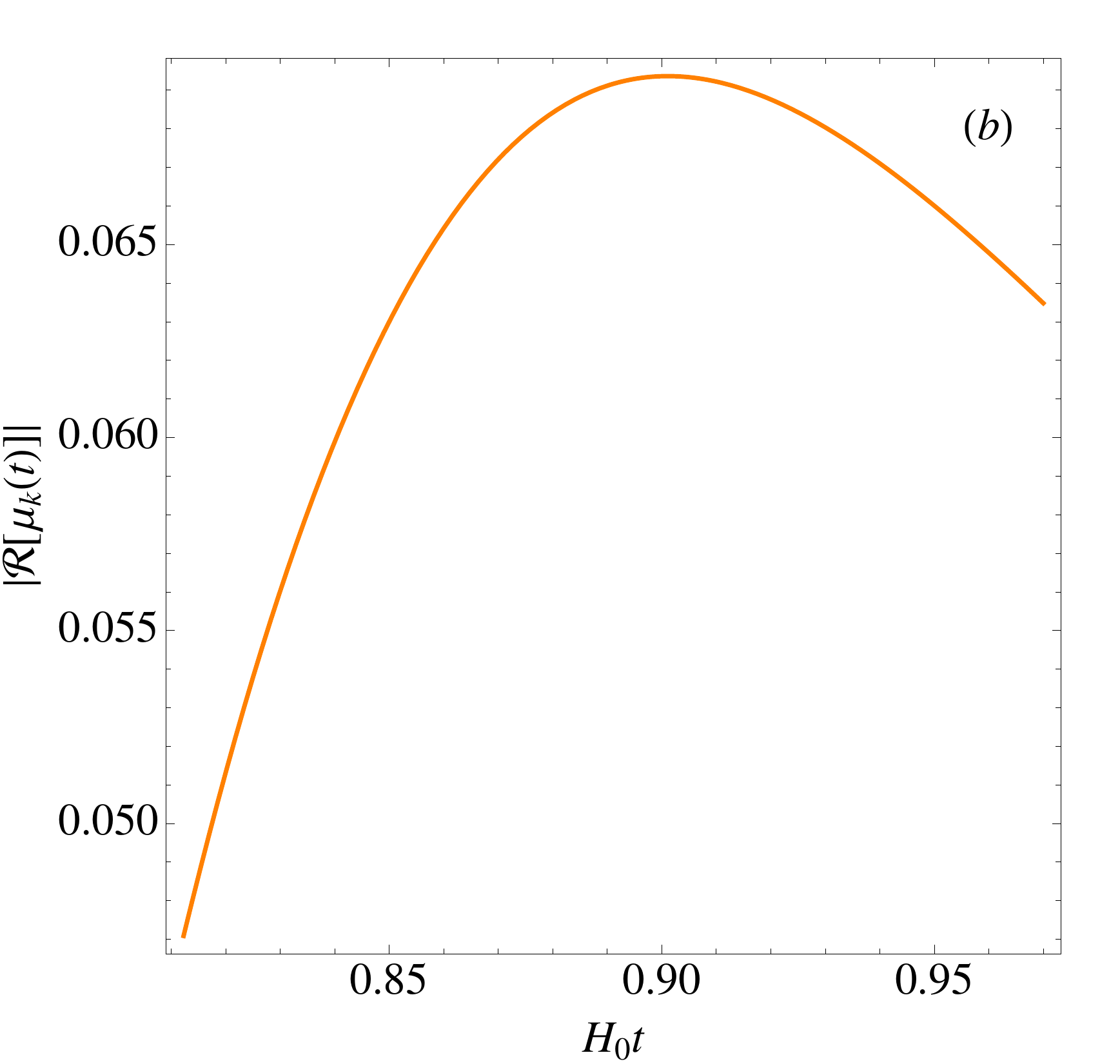} 
\end{array}$
\end{center}
   \caption{In Fig. 4(a), we plot the magnitude of the real part of the Floquet exponent $|\Re(\mu_k)|$ as a function of the physical wavenumber $k_{p}$ and the amplitude of the background field $\vp_{\rm{osc}}$. Yellow corresponds to a larger value than orange, whereas dark red regions have $\Re(\mu_k)=0$. As the Universe expands, the wavenumber of a given mode as well as the amplitude of the background field redshift. As a result, a given mode ``sees" different Floquet exponents as it traverses the Floquet chart along the thin white lines. In Fig. 4(b) we show the Floquet exponent ``seen" by a mode with a comoving wavenumber $k\sim 0.05m$ (along the left-most dashed white line in Fig. 4(a). Modes with $k\gtrsim 0.5m$ pass through many resonance bands in a single oscillation of the homogeneous field and undergo stochastic resonance (see text).}
   \label{fig:Floquet}
\end{figure}
\subsubsection*{Resonance in an expanding universe}
The equation of motion for $\delta\vp_k$ in an expanding universe (still ignoring $\Psi_k$) is
\Beq
\ddot{\delta\vp}_k+3H\dot{\delta\vp}_k+\left[\frac{k^2}{a^2}+U''(\vp)\right]\delta\vp_k=0.
\Eeq
To understand parametric resonance in an expanding background, we make the following identifications: 
\begin{equation}
\begin{aligned}
\label{eq:id}
&k a^{-1}\rightarrow k_p,\\
&\vp_{\rm{osc}} (a)\rightarrow\vp_{\rm{osc}}.
\end{aligned}
\end{equation}
Here $k_p$ is the physical wavenumber and $\vp_{\rm{osc}}(a)$ is the decaying envelope of the oscillating field. This identification \eqref{eq:id} defines a trajectory in the $k_p-\vp_{\rm{osc}}$ plane (white lines in Fig. \ref{fig:Floquet}(a)). The approximate amount of amplification undergone by a given mode is obtained by integrating $|\Re(\mu_k)|$ along the corresponding trajectory in the $k_p-\vp_{\rm{osc}}$ plane. The $|\Re[\mu_k(t)]|$ as seen along such a trajectory (left-most dashed, white line in  Fig. \ref{fig:Floquet}(a)) is shown in Fig. \ref{fig:Floquet}(b). Then, schematically, the evolution envelope of the amplified modes is  (ignoring the oscillatory piece $P_{\pm}(t)$),

\begin{equation}
\label{eq:evolfluc}
\delta\varphi_k(t)\sim \frac{\delta\varphi_k(t_i)}{a^{\gamma}(t)}\exp\left[\int^t d\tau |\Re[\mu_k(\tau)]|\right]
=\frac{\delta\varphi_k(t_i)}{a^{\gamma}}\exp\left[\int^a d\ln \bar{a} \frac{|\Re[\mu_k(\bar{a})]|}{H(\bar{a})} \right]
\end{equation}
where in the second equality we use the scalefactor $a$ as a time co-ordinate and $\gamma=3/2$ when $\vp_{\rm{osc}}\ll M$. 
We assume that the frequency of oscillation $\omega\gg H$. This expression \eqref{eq:evolfluc} is meant to give intuition about the resonant behavior in an expanding universe and should be used with care, especially when multiple bands are involved since the phase of the oscillations can play a role when different bands are traversed. 

When a mode traverses only the first resonance band, the fluctuations get a boost every time the homogeneous field crosses zero. This can be thought of as a burst of particle production. When a large number of bands are traversed within a single oscillation of the homogeneous field, we enter the regime of stochastic resonance. The amplitude of the mode still changes dramatically  at zero crossings of the homogeneous field, however we are no longer guaranteed growth at every such instant. Although over longer time scales the fluctuations grow, at some zero crossings they can also decrease. For our scenario stochastic resonance is seen for modes with $k\gtrsim 0.5m$. A more detailed discussion of these different regimes can be found in \cite{Kofman:1997yn}. 

For our scenario, the modes that grow the fastest are the ones with $k\ll m$ (ones traversing the first band). To get a sense of what is required of the parameters for these modes to grow rapidly, let us now concentrate on the right most expression in \eqref{eq:evolfluc}. If the argument of the exponent is significantly larger than 1, then we will have rapid growth in fluctuations. Using $a_{\textrm{\rm{osc}}}\approx 0.8$, resonance takes place in the logarithmic interval $\Delta \ln a\sim 0.2$. Hence, one gets rapid growth in fluctuations when 
$
|\Re(\mu_k)|/{H}\gtrsim 10.
$
During the oscillatory regime 
$H\approx H_0\sim m \left({M}/{\mpl}\right)$
while from Fig. \ref{fig:Floquet}, the real part of the Floquet exponent seen by a mode in the first band has value of  $|\Re(\mu_k)|\sim 0.1m$. Thus we need $M/\mpl\lesssim 10^{-2}$ for efficient resonance. However as we saw in Sec. \eqref{s:Hom}, in the class of models described by \eqref{eq:potentialAlpha}, consistency with the observed expansion history and the requirement of a few oscillations in the homogeneous field close to today automatically yields $M/\mpl\lesssim(\Delta a)^{3/2}\approx.03$ [Eq. \eqref{eq:alpha}]. Hence in such models, resonance is almost inevitable. 

Importantly, this is a nongravitationally driven growth, driven by the homogeneous, oscillating field $\omega\gg H$. Hence this growth can happen on a time scale which is {\it{significantly shorter}} than $H^{-1}$. 

\subsubsection*{Resonance in an expanding universe including local gravity}
Let us now include the gravitational potentials in the equations of motion and consider resonant phenomenon in the complete coupled system given by Eq. \eqref{eq:PertEOM}. When the field driven resonance is efficient, the gravitational potential $\Psi_k$ does not play a significant role in the field dynamics. The evolution of  $\Psi_k$, on the other hand, is affected significantly by the resonant growth in $\delta\vp_k$ since the quintessence field dominates the energy density. We stress that $\Psi_k$ can only be ignored in the evolution of $\delta\vp_k$ for modes undergoing strong resonance when the self interactions dominate over the gravitational one. In particular, as we have seen in the previous sections $\delta\vp_k$ just before resonance is determined by $\Psi_k$ and as a result ignoring $\Psi_k$ before resonance is not justified. In addition, the gravitational potential includes contributions from the WIMP dark matter overdensity via the constraint equations. These two considerations make the analysis of Eq. \eqref{eq:PertEOM} in the resonance regime nontrivial and we have to rely on numerical solutions. A more detailed {\it analytical} analysis of resonance in an expanding universe including the effects of the gravitational potential will be pursued elsewhere. Below we discuss the numerical solutions during resonance in a bit more detail.

Typical, rapidly growing solutions for $\delta\vp_k$ and $\Psi_k$ with $k\approx 0.05m$ are shown in Fig. \ref{fig:ResGrowth}(a) and (b), respectively. Note the rapid, nearly exponential growth of $\Psi_k$ and $\delta\vp_k$ after $a_{\textrm{\rm{osc}}}\approx 0.8$ (vertical, dashed line). The thin, black lines show the evolution of the same modes in $\Lambda$CDM. The growth in the gravitational potential is somewhat delayed compared to the field. This is because the energy density in the field has to first become comparable to that of dark matter. Before this happens, the gravitational potential does not experience scale dependent growth (though it still responds to the changing expansion history).

 \begin{figure}[H]
\centering
\includegraphics[width=4.5in]{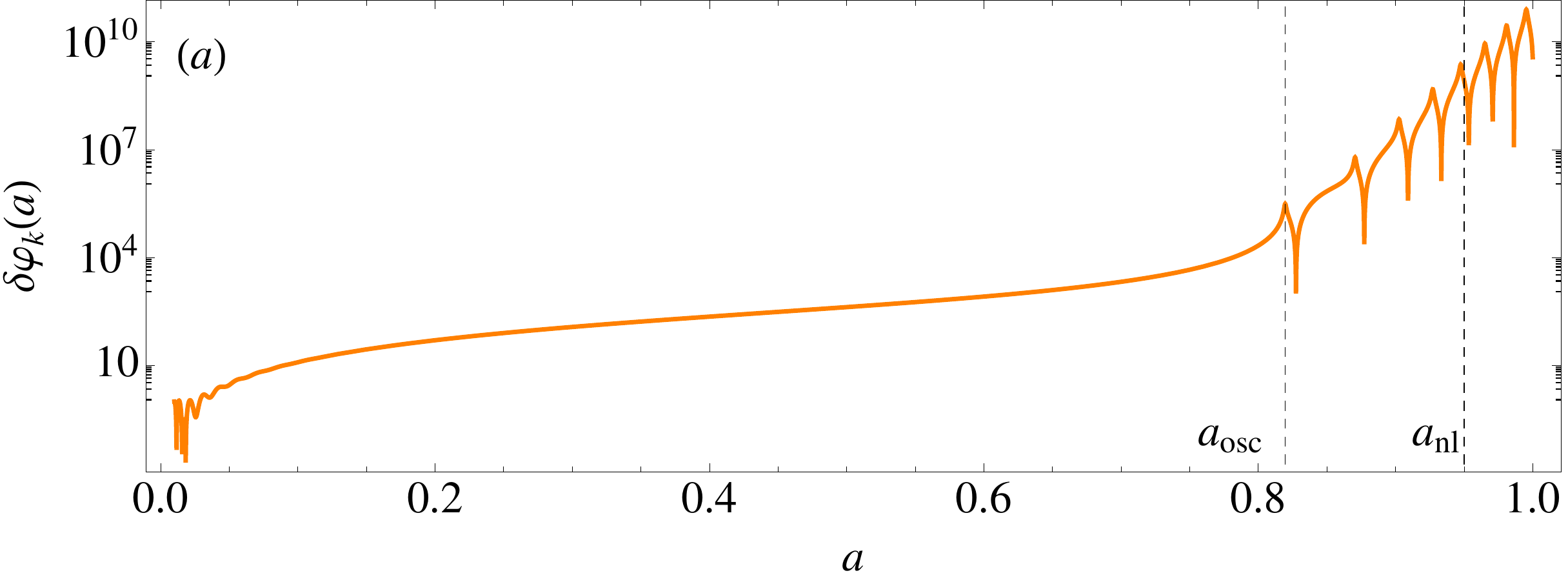}\\
\includegraphics[width=4.5in]{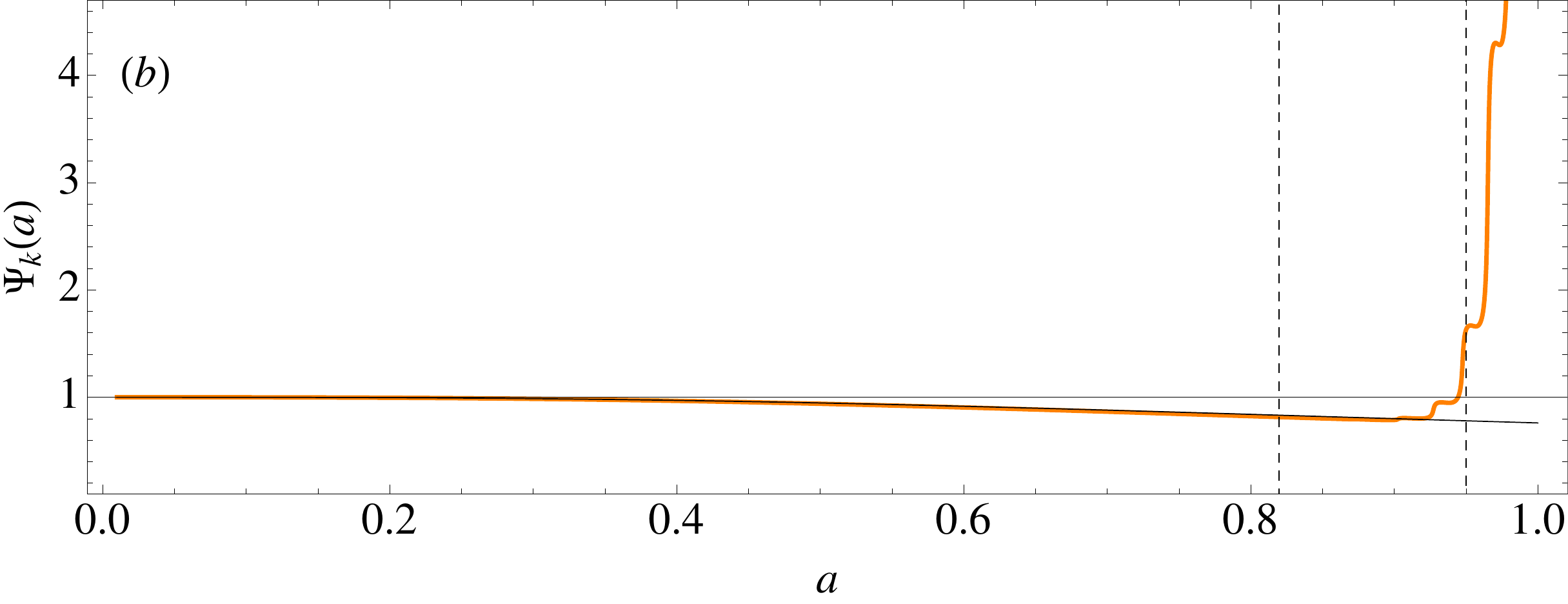}\\
 \includegraphics[width=4.5in]{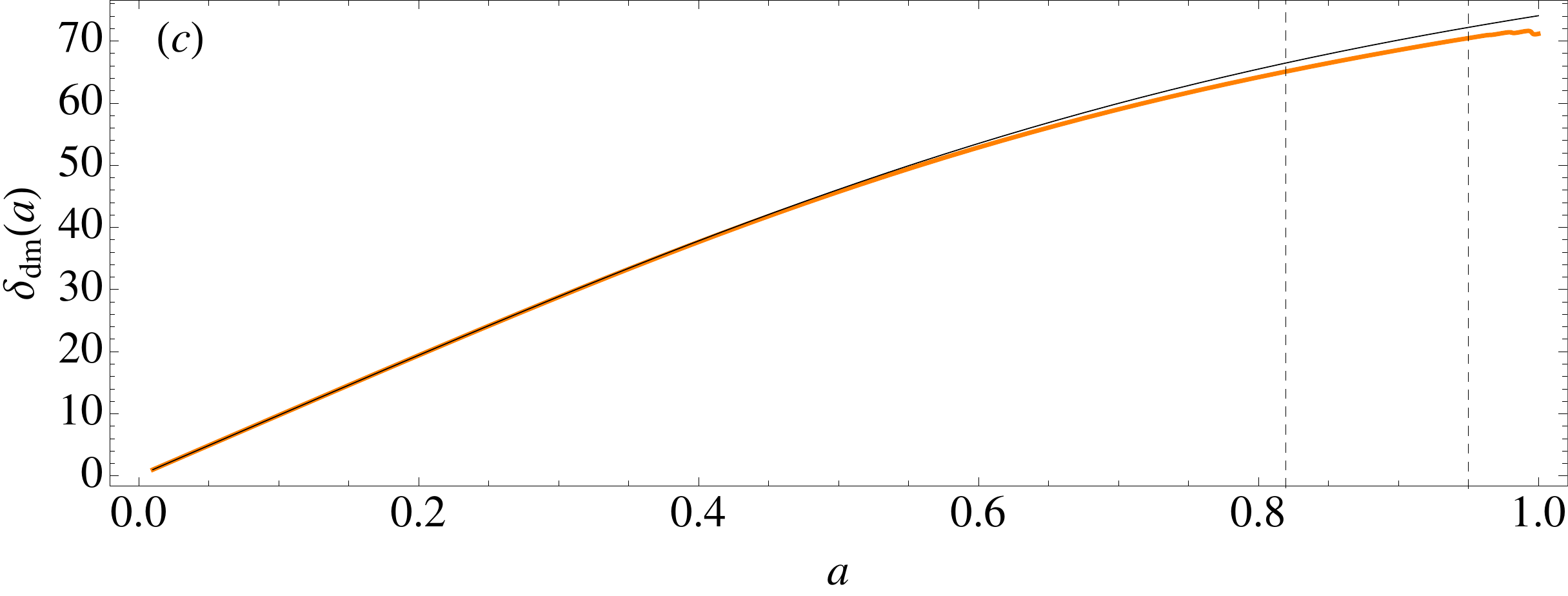}  
   \caption{(a) Evolution of $\delta\vp_k$  with $k\sim 0.05m\approx 0.01\rm{Mpc}^{-1}$. During matter domination, $\delta\vp_k\propto a^2$ and grows relatively slowly. However, after $\aosc$, the mode grows exponentially fast due to parametric resonance. At $\anl$ the fluctuations become nonlinear and linear evolution is no longer applicable. (b) Evolution of the scalar gravitational perturbation ($k\sim 0.05m$). Before resonance, the evolution is similar to that expected in usual slow roll quintessence (or $\Lambda$CDM, thin black line) model. It is constant deep in the matter dominated era and starts decaying as quintessence takes over. Unlike slow roll quintessence, after $\aosc$ the potential grows rapidly until $\anl$. This leads to a scale dependent signal in observations that are sensitive to the gravitational potential.  (c) The evolution of the WIMP overdensity is not significantly affected by the resonant growth. The departure from $\Lambda$CDM in this case is almost entirely due to the slight deviations in the expansion history. The normalization is set to one for all the above modes at $a=a_i$. }
   \label{fig:ResGrowth}
\end{figure}

So far we have ignored discussing the evolution of the WIMP overdensity since it is determined from the constraint Eq. \eqref{eq:dmConstraint}. Naively one might expect to see scale dependent departures in the behavior of $\delta_{\rm{dm}}$. However, despite the rapid growth in the gravitational potential after $\aosc$, $\delta_{\rm{dm}}$ does not deviate significantly from its $\Lambda$CDM counterpart (see \ref{fig:ResGrowth}(c)) in the linear regime. Although difficult to see from the constraint Eq. \eqref{eq:dmConstraint}, this behavior can be understood by considering the conservation equation for the WIMPs,  
\Beq
\ddot{\delta}_{\rm{dm}}+2H\dot{\delta}_{\rm{dm}}=-\frac{k^2}{a^2}\Psi_k+3\left(2H\dot{\Psi}_k+\ddot{\Psi}_k\right).
\Eeq
Heuristically, we see that on subhorizon scales, $\delta_{\rm{dm}}$ is obtained from a double time integral of the potential $\Psi_k$. This delays the response of the $\delta_{\rm{dm}}$ to the $\Psi_k$. The small departure of  $\delta_{\rm{dm}}$ from its $\Lambda$CDM counterpart can be accounted for by the difference in expansion history between the quintessence and the $\Lambda$CDM models. In particular, since the Hubble parameter in our quintessence model is always slightly larger than its $\Lambda$CDM counterpart (see Sec. \ref{s:Hom}), the growth of $\delta_{\rm{dm}}$ in the quintessence model is slightly suppressed.  

From the above discussion we see that the linearized fluctuations in the field and the gravitational potential grow rapidly. Eventually, the field fluctuations will become nonlinear and we cannot trust the linearized treatment. Hence it becomes important to understand, at least qualitatively, when the field becomes nonlinear as well as what happens thereafter. Although we discuss the nonlinearity of the scalar field fluctuations, we do not include the usual nonlinearity in $\delta_{\rm{dm}}$ at late times. This nonlinearity in $\delta_{\rm{dm}}$ is of course well studied, but to include it would take us too far beyond the scope of this paper. 
 
 \subsubsection{Nonlinearity and oscillon formation}
We cannot ignore the nonlinearity of the fluctuations when $U'(\vp)\sim U''(\vp)\delta\vp$. In the oscillatory regime this happen at the scale factor $\anl$ when
\Beq
\label{eq:NLCondition}
\Delta_{\delta\vp}(k,\anl)\sim\vp_{\rm{osc}}(\anl).
\Eeq
where $\Delta^2_{\delta\vp}(k,a)=k^3P_{\delta\vp}(k,a)/2\pi^2$ with $\langle\delta\vp_k\delta\vp_{k'}\rangle\equiv (2\pi)^3P_{\delta\vp}(k)\delta^3(k-k')$ is the power spectrum of the field fluctuations (see Fig. \ref{fig:SpectraField}(a)).  Heuristically, the left hand side characterizes the mean square fluctuations over a spatial region of size $L$, i.e.
\Beq
\langle\delta\vp^2\rangle^{1/2}_{L}=\left[\Delta_{\delta\vp}(k,a)\right]_{k\sim L^{-1}}
\Eeq
In  Fig. \ref{fig:SpectraField}(b) the orange curves show the evolution of the (absolute value of) oscillatory homogeneous field and its envelope. The black curves show the maximum value of $\Delta_{\delta\vp}(k,a)$ as a function of $a$ and its envelope. The location where the two curves intersect $\anl\approx0.95$ is taken as the point beyond which the linearized equations cannot be trusted. In Fig. \ref{fig:SpectraField}(a) we show $\Delta_{\delta\vp}(k,\anl)$ (solid, orange curve) and $\Delta_{\delta\vp}(k,a_i)$. As mentioned in Sec. \ref{ss:MDPert}, the normalization of $\Delta_{\delta\vp}(k,a_i)$ is set by $\Delta_{\Psi}(k,a_i)$, which we take from WMAP 7 \cite{Komatsu:2010fb}($\Delta^2_\mathcal{R}=(9/25)\Delta^2_{\Psi}=2.42\times 10^{-9}$ at $k=0.002\;\rm{Mpc}^{-1}$ and $n_s=0.966$). We also note that although $k\sim 0.05m \approx 0.01\;\rm{Mpc}^{-1}$ is where we see the maximum deviation, resonance significantly enhances fluctuations for larger $k$ as well, albeit through stochastic resonance. 

In order to calculate the effect of the rapid growth of the gravitational potential on observables, we need its evolution until today, ie. $a\approx 1$. However, as discussed above, the nonlinearities of the field do not allow us to compute it for $a>\anl=0.95$. To remedy this, for the linearized calculation, we take a conservative approach and ``freeze" the value of the potential at $\anl$, setting $\Psi_k(a)=\Psi_k(\anl)$ for $\anl\le a\le 1$. More realistically, the potential will evolve further in a scale-dependent manner as the scalar field perturbations undergo nonlinear evolution. Fragmentation of the scalar field [discussed below] will enhance the perturbations, at least on smaller scales [we provide some simple estimates of the ISW effect due to this nonlinear evolution in Appendix B]. Along with the break-down of linearized equations, we also note that after $\anl$, large gradients in the scale field fluctuations will source anisotropic stress, violating our assumption of $\Psi= \Phi$. 
\begin{figure}[t] 
\begin{center}$
\begin{array}{cc}
   \includegraphics[width=3.1in]{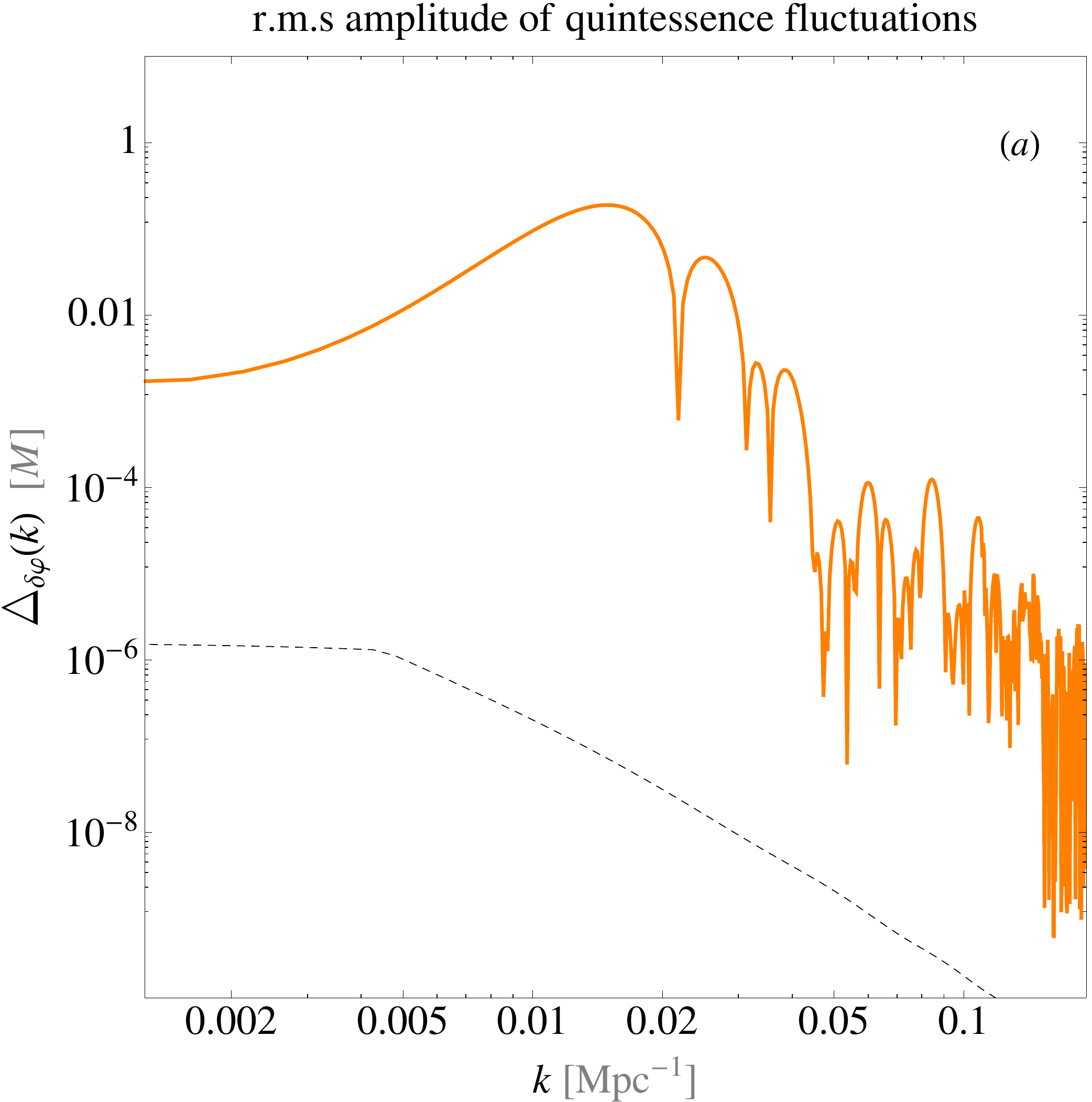}&
 \includegraphics[width=3.in]{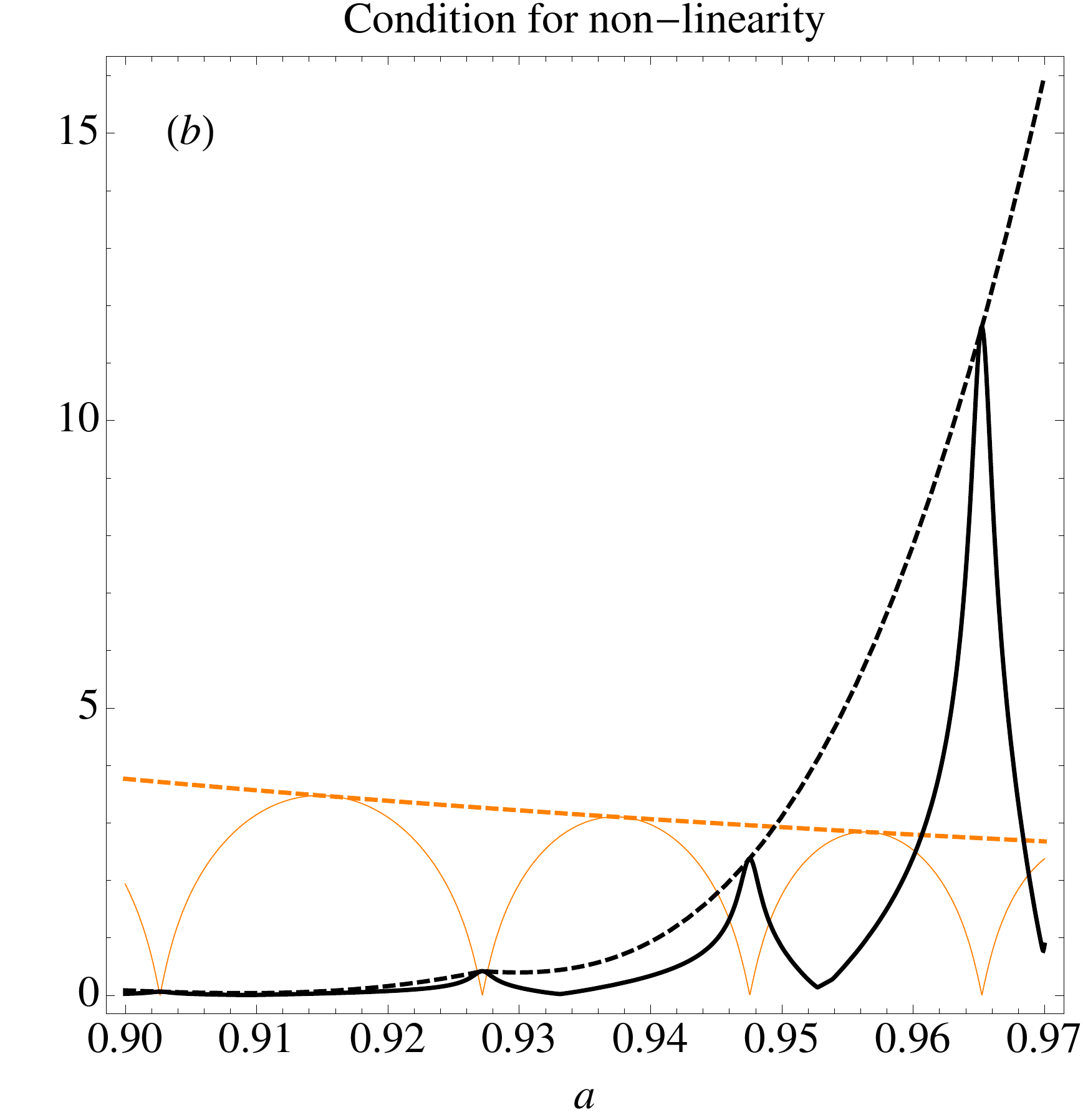} 
\end{array}$ 
\end{center}
 
   \caption{(a) Fluctuations in quintessence $\langle\delta\vp^2\rangle^{1/2}_{L\sim k^{-1}}= \Delta_{\vp}(k,a)=\sqrt{k^3P_{\delta\vp}(k,a)/2\pi^2}$ at $\anl=0.95$, whereas the dashed curve is $10^3\times$ the initial conditions at $a_i=0.01$.  Note the large bump due to resonant growth seen in the $\delta\vp$ field at $k\approx0.05m\approx0.01\,\rm{Mpc}^{-1}$. The higher $k$ regions undergo stochastic resonance. (b) To determine when the fluctuations become nonlinear ($\anl$), we compare the rms fluctuations to the background field. The orange curves denote the homogeneous field whereas the black curves denote the maximum value of $\Delta_{\delta\vp}(k,a)$ at each $a$. Note the enhancement at each zero crossing. Once the field fluctuations become nonlinear, different $k$ modes couple and the field fragments into localized, long-lived energy density configurations (oscillons).} 
   \label{fig:SpectraField}
\end{figure}

Although in this paper we do not pursue nonlinear evolution of the scalar field fluctuations, below we digress slightly and point out some of the interesting phenomenology that results. As nonlinearity sets in, the evolution of different $k$ modes becomes coupled, and back-reaction from the perturbations curtails the resonant growth of fluctuations. The homogeneous field then fragments rapidly. For the type of potentials considered here (those having a quadratic minimum and a shallower than quadratic form away from the minimum), most of the energy density can eventually end up in localized, oscillatory, long-lived configurations of the field called oscillons (e.g. \cite{Bogolyubsky:1976yu,Gleiser:1993pt, Amin:2010jq}). The central density of these oscillons can be greater than the background energy density and their sizes are of order a few $m^{-1}\approx 10^{-3}H_0^{-1}\approx 4\, \rm{Mpc}$. Although oscillons radiate energy through scalar radiation (e.g. \cite{Segur:1987mg, Fodor:2008du, Gleiser:2008ty, Hertzberg:2010yz}), our limited radial simulations of individual oscillons reveal a lifetime of $\sim10^6m^{-1}\sim10^3H_0^{-1}$ for an oscillon with width of a few $m^{-1}$ and field amplitude of order $M$. This makes them effectively stable compared to current cosmological time scales. Unlike the usual oscillons (e.g.\cite{Bogolyubsky:1976yu,Gleiser:1993pt, Gleiser:2008ty,Amin:2010jq}), which have a relatively stationary energy density profile, the oscillons in our model have an energy density that breathes in and out (also see \cite{Amin:2011hj}) at about twice the field oscillation frequency. 

This phenomenon of oscillon production has been studied in the context of the early Universe, in particular, at the end of a similar scalar field driven inflationary period (e.g. \cite{Gleiser:2011xj, Amin:2010dc,Amin:2010xe,Amin:2011hj}). Here, we point out that it will likely happen at the end of the current period of cosmic acceleration as well. As with the early Universe studies, we expect the oscillons to dominate the energy density. However, unlike the early Universe, these oscillons have sizes which make them astrophysically accessible with novel signatures such as the existence of dark, low redshift clusters with time dependent configurations made of the quintessence field. We stress that it is not guaranteed that the Universe will become oscillon dominated before today. Although for the fiducial parameters, the field fluctuations will become nonlinear before today, it can take some time for the nonlinear field to enter a state where it is dominated by oscillons. This time can be longer than the time between $\anl$ and today. Quantifying this process, requires simulating the full nonlinear dynamics of the quintessence field, including nonlinearities of the WIMP overdensity, which is beyond the scope of this paper. Given the similarity of the potential with \cite{Amin:2011hj}, we expect many of the qualitative results to carry over. However, to perform a reliable calculation to be compared with observations, one cannot completely ignore the gravitational perturbations, which makes the simulations somewhat more challenging. 
\begin{figure}[t] 
\begin{center}$
\begin{array}{cc}
 \includegraphics[width=2.9in]{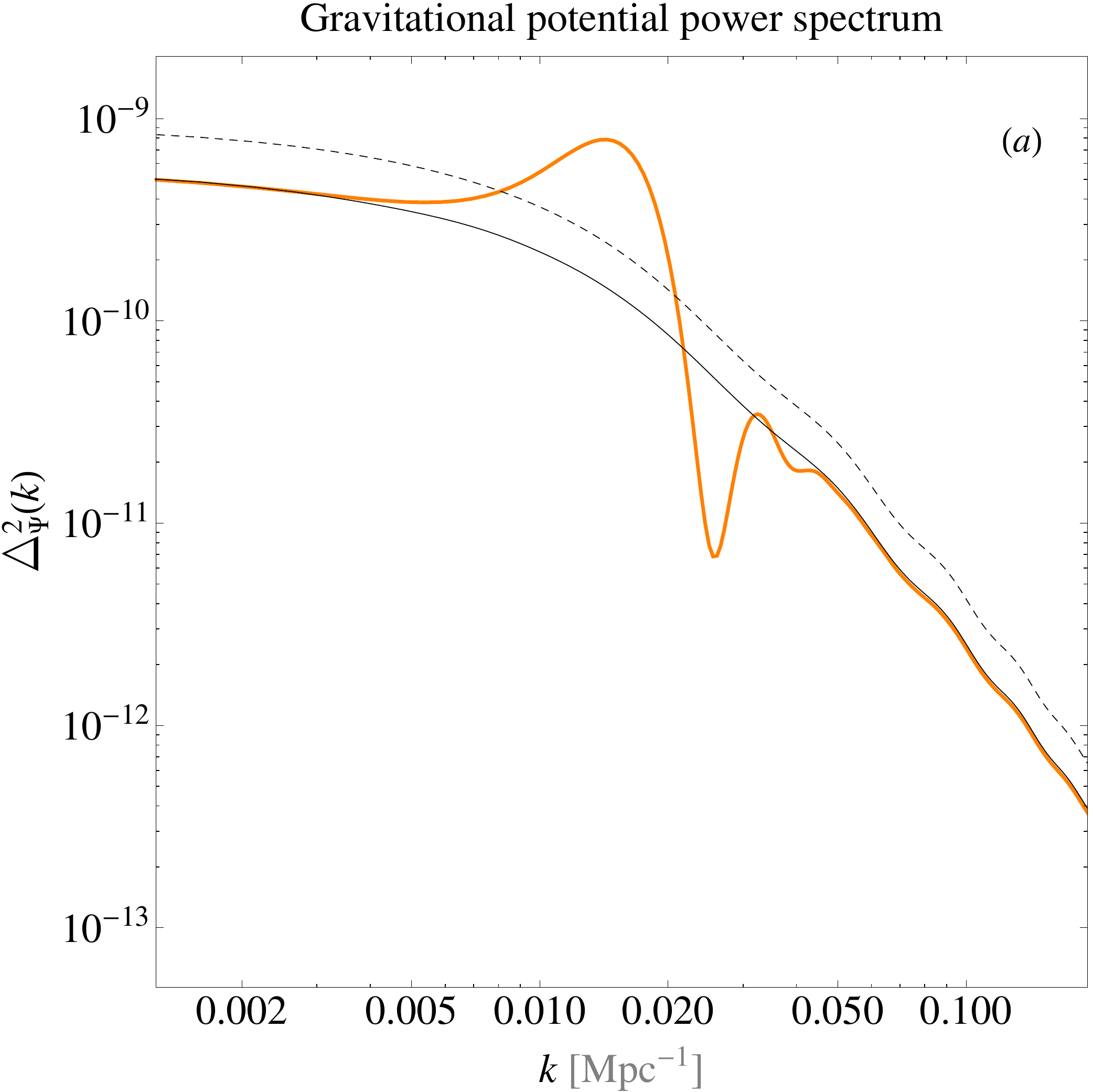} &
 \includegraphics[width=3.1in]{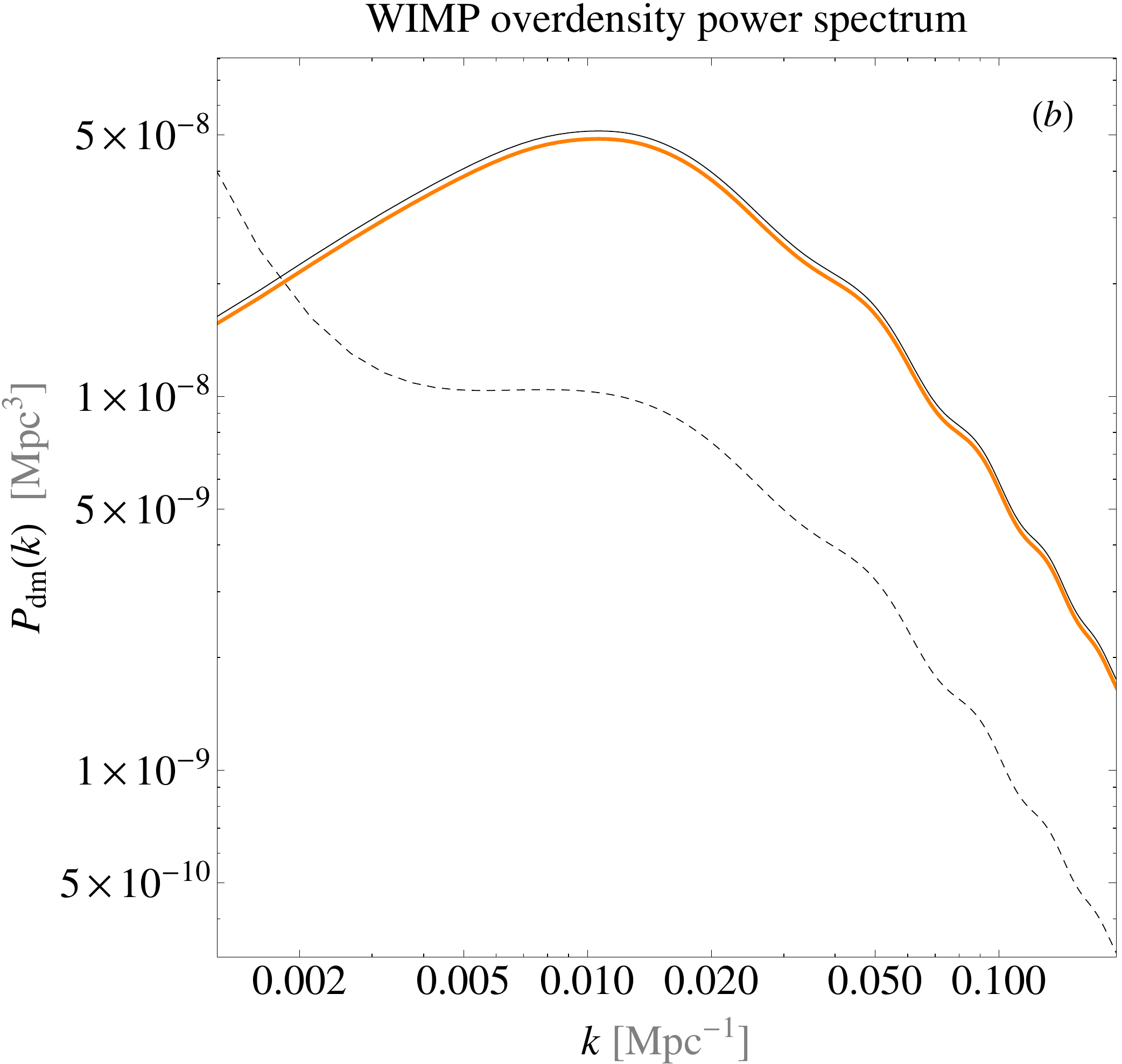} 
\end{array}$ 
\end{center}
   \caption{(a) Power spectrum of the gravitation potential (orange). The dashed line denotes the initial conditions at $a_i=0.01$ whereas the thin black line shows the $\Lambda$CDM power spectrum today. Notice the scale dependent growth of the gravitational potential for $k\sim 0.05m=0.01\rm{Mpc}^{-1}$, aside from which the potential decays in a manner similar to its $\Lambda$CDM counterpart. (b) The WIMP power spectrum (orange line) does not show any features of the scale dependent growth, since the WIMPs do not have enough time to respond to the change in the gravitational potential. The difference from the $\Lambda$CDM counterpart (black line) is primarily due to the difference in expansion history between the quintessence and $\Lambda$CDM models. The dashed line represents $10^3\times P_{{\rm{dm}}}(k,a_i)$. Note that since we are using $\delta_{\rm{dm}}$ in the Newtonian gauge, $P_{\delta_{\rm{dm}}}(k,a_i)$ shows an ``upward turn" at small $k$ values.} 
   \label{fig:Spectra}
\end{figure}

\section{Power spectra and observables}
\label{s:Obs}
With our understanding of the evolution of individual modes and the validity of the linearized equations, we now compute the power spectra 
of the gravitational potential $\Delta^2_{\Psi}(k,a)\equiv k^3P_{\Psi}(k,a)/2\pi^2$ and the WIMP overdensity $P_{\delta_{\rm{dm}}}(k,a)$ and their impact on observables such as galaxy clustering, lensing and the CMB.

The calculated power spectra $\Delta^2_{\Psi}(k,\anl)$ and $P_{\delta_{\rm{dm}}}(k,\anl)$ (with $\anl=0.95$) are shown in Fig. \ref{fig:Spectra} (thick, orange lines). The thin, black lines show the corresponding power spectra for the $\Lambda$CDM cosmology. The gray, dashed lines show the power spectra at $a_i=10^{-2}$. Note that since we are using $\delta_{\rm{dm}}$ in the Newtonian gauge, $P_{\delta_{\rm{dm}}}(k,a_i)$ shows an ``upward turn" at small $k$ values. By construction, the initial conditions in the power spectra $\Delta_{\Psi}(k,a_i)$ and  $P_{{\rm{dm}}}(k,a_i)$ in  our quintessence model and $\Lambda$CDM agree. 
The scale dependent departures in $\Delta_{\Psi}(k,a)$ arise after $\aosc\approx 0.8$. As discussed previously, the enhancement can be trusted until $\anl=0.95$, after which we freeze the potential until $a=1$. The wavenumber where the maximum departures are seen is $k\sim0.05m\approx0.01\,\rm{Mpc}^{-1}$. For $k\ll H_0$ and $k\gtrsim m$, the final spectrum closely resembles the $\Lambda$CDM case. For fixed $\aosc$, we have checked that the magnitude of departure from $\Lambda$CDM increases with increasing $m$. This is consistent with the notion larger $m$ corresponds to a larger $\mpl/M$ [see Eq. \eqref{eq:constraints}] leading to more efficient resonance. As discussed for single modes, the potential $\Delta_{\Psi}(k,a)$ show dramatic, scale dependent departures from $\Lambda$CDM, but the WIMP overdensity power spectrum does not. 

Assuming a model for the bias, the WIMP power spectrum can be probed by measuring the two point correlation function of galaxies. Given the lack of scale dependence, and the small difference ($\lesssim 6\%$) with $\Lambda$CDM, we expect the departures from $\Lambda$CDM to be difficult to detect. \footnote{Note that this is reasonable since, like the WIMPs, galaxies will not have enough time to respond to the late time scale dependent quintessence fluctuations.} The fluctuations in quintessence are not directly observable except through their gravitational imprint.  This leaves us with the gravitational potential as the key probe to look for departures from $\Lambda$CDM. The rapidly changing potential in the quintessence model and the large scale dependent departures at low redshifts could affect the low multipoles of the CMB temperature and weak lensing power spectrum. With this in mind, we now calculate the weak lensing (convergence) power spectra and the CMB temperature anisotropy. We remind the reader that the differences that arise here are in spite of having made the expansion history consistent with current observations. An expansion history and galaxy power spectrum consistent with \lcdm but scale dependent deviations in the lensing and CMB power spectra provides a unique probe of a late time transition in the quintessence field.

\subsection{Integrated Sachs-Wolfe effect}

When CMB photons travel through an evolving gravitational potential their energy changes (the integrated Sachs-Wolfe (ISW) effect \cite{Sachs:1967er}). Since the rapid growth in the quintessence fluctuations induces large changes in the potential, we expect imprints of the late-time quintessence transition on the CMB power spectrum through the ISW effect. 

The angular power spectrum of the CMB temperature can be written as \cite{Ma:1995ey}
\Beq
C_l=4\pi\int d\ln k\,\, D^2_l(k) \Delta^{2}_{\Psi}(k)|_{\rm{prim}}
\Eeq
where $\Delta^{2}_{\Psi}(k)|_{\rm{prim}}$ is the primordial power spectrum. In the above, $D_l(k)\equiv \Delta_l(k)/\Psi^{\rm{prim}}_k$, where $\Delta_l(k)$ on large angular scales (under the assumptions of adiabatic initial conditions and instantaneous recombination) is given by,
\Beq
\Delta_l(k)
&=\Delta^{SW}_l(k)+\Delta^{ISW}_l(k)\\
&\approx \left[\frac{1}{3}j_l(k\chi_e)+\frac{2}{3}\frac{k}{a_eH_e}j_l'(k\chi_e)\right]\Psi_k(a_e)+\int_{a_e}^{1}da\,\,j_l(k\chi_a)\partial_a\left[\Psi_k(a)+\Phi_k(a)\right].
\Eeq 
In the above expression, the subscript ``e" stands for emission (ie. surface of last scattering), $\chi_a=\int_a^1da/a^2H(a)$ is the comoving distance, and $j_l(x)$ are spherical Bessel functions. The second term is the ISW term, which gets large if the potential evolves significantly between last scattering and today. In \lcdm the ISW term gets a small contribution just after recombination since we are not quite matter dominated at that time. But, the main contribution on the scales of interest comes from late times as the universe starts becoming $\Lambda$ dominated and the potentials start decaying. The same effect is also present in the quintessence model. However, there is an additional contribution from the late-time rapid growth of the potential. This growth occurs in spurts [see Fig. \ref{fig:ResGrowth}(b)] and we can approximate $\partial_a\Psi_k$ as a Dirac-delta function at the location of the jumps in the gravitational potential. As a result, the contribution of each jump to the ISW term can be easily evaluated:
 \Beq
{}^{(j)}\Delta^{\rm{ISW}}_l(k)\approx2 j_l(k\chi_{a_j})\Delta \Psi_k(a_j),
\Eeq
where $a_j$ is the scale factor where the potential jumps. For a jump of order a few $\Psi_k$, the ``jump" term is larger than the smooth ISW contribution in $\Lambda$CDM. Note that there can be multiple jumps, both positive and negative, depending on the $k$ mode in question. Also note that apart from the magnitude of the jump, the location $a_j$ also plays a role by fixing the argument of the Bessel function.

We expect that this growth affects the low multipole  moments of the CMB since the changes in the potential occur very recently.  More specifically, the multipole range where we expect deviations is $l\lesssim k \chi_{a_j}$, where $k\sim 0.05 m\approx 0.01\rm{Mpc}^{-1}$.  For the fiducial parameters chosen here, this yields $l\lesssim$ few. Note that along with $k$, it is the smallness of $\chi_{a_j}$ that limits the effect to large angular scales. 

In order to quantitatively calculate the effect of the quintessence perturbations on the CMB power spectrum, we use the $\Psi_k(a)$ and $\Phi_k(a)$ evaluated from CMBFAST \cite{Seljak:1996is} for $a<a_i=0.01$. This captures the contributions from early ISW as well as early anisotropic stress (which we ignore at late times). For $a>a_i$ we use $\Psi_k(a)$ computed from our own independent code for \lcdm and the quintessence model.  As discussed before, for the quintessence case we set $\partial_a\Psi_k=0$ for $a>\anl$. With this entire solution at hand (for $a_e<a<1$) we compute the $\Delta_l(k)$ and $C_l$.  The ratio of the angular CMB power spectra for the quintessence and \lcdm models is shown in Fig. \ref{fig:SpectraObs}(a). Note that this is large enough to potentially rule out this choice of parameters. Thus, we see that in spite of the effect being on large angular scales, ISW provides an excellent probe for constraining the considered transition in quintessence. 
\begin{figure}[t] 
\begin{center}$
\begin{array}{cc}
\includegraphics[width=2.75in]{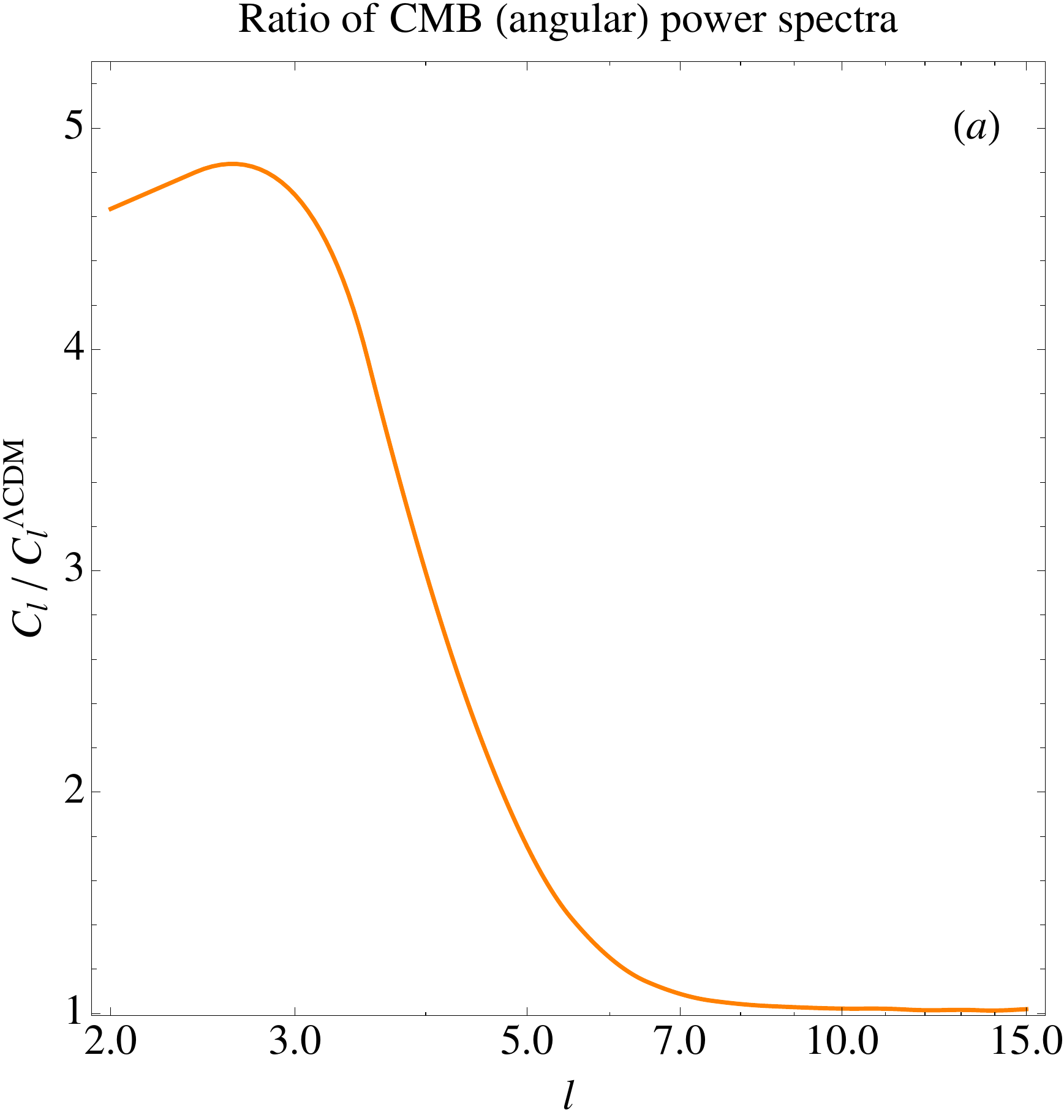} &
\includegraphics[width=2.8in]{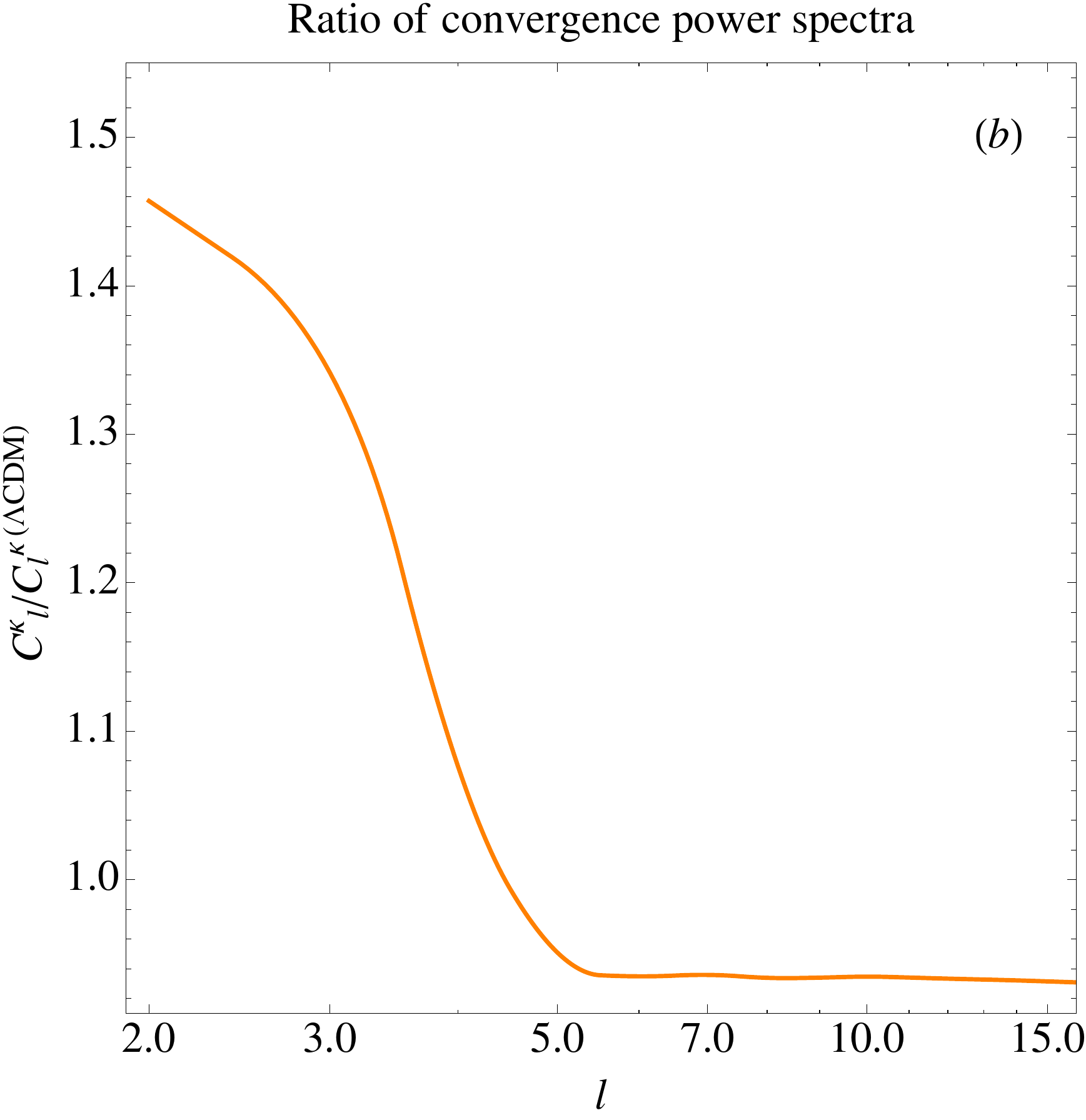}
\end{array}$ 
\end{center}
 
   \caption{(a) Ratio of CMB temperature power spectrum at low multipoles between our quintessence and \lcdm models. The large difference is due to the rapidly evolving gravitational potential in the quintessence model after $\aosc$ (b) Ratio of the lensing (convergence) power spectrum at low multipoles between our quintessence and \lcdm models. The departures are restricted to low $l$ multipoles due to the proximity of the transition in the quintessence field responsible for the rapid growth in the gravitational potential. For lensing, the model does not agree with \lcdm at large $l$ due to the different expansion histories.}
   \label{fig:SpectraObs}
\end{figure}
One can change parameters, for example $\aosc$ or $m$ (equivalently $M$) to get the ISW effect small enough so that it is consistent with observations. One has to first make sure that such changes are still consistent with the expansion history (see discussion in Sec. \ref{s:Hom}).  Increasing $m$ or $\aosc$ or varying $m$ and $\aosc$ in opposite directions can lead to an expansion history consistent with observations. Let us now look at perturbations. In general, for a fixed $\aosc$, as we increase $m$ (equivalently decrease $M$), the fluctuations grow more rapidly, shifting the  time when the potentials grow rapidly to smaller $a$ values. One might expect that this will make the ISW contribution even larger. However, one finds that the effect on $\Delta_l(k)$ is not quite as simple. First the change in growth of potential is quite sensitive to the number of zero crossing of the homogeneous field. In addition, changing parameters changes the wavenumbers, which grow the fastest and the time when there is a rapid growth in the potential. Heuristically, this affects the argument of the $k\chi$ of the nonmonotonic spherical Bessel function in $\Delta_l(k)$. Similar considerations apply to changing $\aosc$. As a result, it is somewhat difficult to apriori predict which combination of parameters (consistent with the observed expansion history) will also yield an acceptable ISW term. Indeed we find that $m\approx 2\times 10^3H_0$ and $\aosc\approx 0.82$ is entirely consistent with observations of the expansion history as well as the CMB.  We also find regions of parameter space with smaller values of $m$ but larger values of $\aosc$ compared to the fiducial model consistent with observations. A full sweep of parameters (including $\alpha$) to determine regions consistent with observations is beyond the scope of this paper, but is certainly worth pursuing. 

Before we move on to the calculation of the lensing power spectrum, let us briefly comment on the additional ISW that can result from nonlinear field evolution after $\anl$, ignored so far in the linearized treatment. As noted in Sec. 4.2.3, the rapid growth of the linearized scalar field perturbations is curtailed once the field perturbations become nonlinear (around $\anl$). However, there will be further nonlinear evolution and rapid  fragmentation of the field. To get an estimate of the resulting ISW effect, consider a spherical perturbation with mass $\mathcal{M}$ and radius $R$, collapsing at speed $v$. Such a perturbation yields a temperature decrement of order $\sim G\mathcal{M}v/R$. For a sphere with an initial radius $\Rnl\sim \knl^{-1}$ (with $\knl\sim 0.05m$), an initial density comparable to the background energy density and $v\sim 0.05$ (the approximate group velocity of a perturbation with wavenumber $\knl$), the decrement can be $\sim {\rm{few}}\times 10^{-5}$ for our fiducial set of parameters ($m\sim 10^{3}H_0$). The signal is qualitatively similar to the Sunyaev-Zel'dovich temperature decrement from galaxy clusters \cite{Marriage:2010cp}. However, unlike $\lesssim \rm{arcmin}$ angular scale of the SZ decrement, here, the angular scale is $\sim 30\, {\rm{degrees}}$. This simple estimate shows that it might be possible to get additional constraints on the quintessence transition from the ISW effect due to the nonlinear field evolution. Individual oscillons, modeled by density configurations with time dependent radii $\sim {\rm{few}} \times m^{-1}$ can also yield an additional ISW signal on smaller scales. We discuss a toy model for estimating the ISW signal from such nonlinear perturbations in Appendix B. However, we again caution the reader that these numbers should be checked with input from detailed simulations of the nonlinear field dynamics including gravity.

\subsection{Weak lensing}
Observational constraints from weak lensing observations are often presented in terms of the weak lensing convergence (angular) power spectrum \cite{Kaiser:1996tp, Hu:1999ek},

\begin{equation}
C^{\kappa}_l=8\pi^2\int^{\chi_e}_0 d\chi W^2(\chi)\frac{l}{\chi}\Delta^2_{\Psi}\left(k=l\chi^{-1},\chi\right)
\end{equation}

\noindent where:

\begin{equation}
W(\chi)=\int^{\chi_e}_\chi d\chi'\frac{\chi'-\chi}{\chi'}\eta(\chi'),
\end{equation}

\noindent $\Delta^2_{\Psi}(k,\chi)$ is the power spectrum of the gravitational potential and $\eta(\chi)$ is the radial distribution of sources, normalized to $\int\eta(\chi)d\chi=1$. We use the source distribution,

\begin{equation}
\eta(z)\propto z^2\exp[-(1.41z/z_{\rm{med}})^{1.5}]
\end{equation}  

\noindent with $z_{\rm{med}}=1.26$. This distribution approximates the galaxy redshift distribution of the COSMOS survey\cite{Massey:2007gh}.  As with the ISW effect, because of the closeness of the transition to the present day, we expect the signal to be largest at low multipoles $l\lesssim k \chi({\anl})\sim$ few. The proximity of the transition also picks out sources at approximately $2\chi({\anl})$. As a result as $\anl\rightarrow 1(\chi\rightarrow 0)$, we run out of sources to be lensed. Thus in general, for a fixed enhancement of the gravitational potential, we expect the largest lensing signal to arise from the smallest $\anl$  still consistent with the expansion history.  
 
The ratio between the weak lensing convergence power spectrum for the quintessence model and \lcdm is shown in Fig. \ref{fig:SpectraObs}(b). The errors associated with measuring the convergence power spectrum is given by \cite{Kaiser:1996tp}:

\begin{equation}
\triangle C^{\kappa}_l=\sqrt{\frac{2}{(2l+1)f_{\rm{sky}}}}\left(C^{\kappa}_l+\frac{\langle\gamma^2_{\rm{int}}\rangle}{\bar{n}}\right)
\end{equation}

\noindent where $f_{\rm{sky}}$ is the fraction of the sky covered by the survey, $\bar{n}$ is the measured galaxy number density, and $\langle\gamma^2_{\rm{int}}\rangle^{1/2}\approx .4$ is the galaxy intrinsic rms shear in one component. Using survey parameters characteristic of the Large Synoptic Survey Telescope (LSST) ($f_{\rm{sky}}\approx .48$, $\bar{n}\approx 5.9\times10^8\rm{sr}^{-1}$) \cite{Ivezic:2008fe}, we find $\triangle C^{\kappa}_l/C^{\kappa}_l\approx \{.9,.26\}$ for $l=\{2,30\}$. Hence this particular quintessence model cannot be constrained using LSST measurements of the weak lensing power spectrum. The weak lensing convergence power spectrum for the quintessence model does not asymptote to the \lcdm prediction at higher $l$ because there is a difference in expansion history between the two models. Even this deviation would be difficult to see using LSST since the minimum $\triangle C^{\kappa}_l/C^{\kappa}_l \approx 0.1$ at $l\sim 300$.

Before moving on to our conclusions, we note that there are qualitative degeneracies between observational signatures (in particular, ISW at low multipoles) predicted by our scenario and those of other models, for example, interacting dark energy  models with significant clustering (see \cite{Amendola:2007yx,Pettorino:2010bv, Baldi:2011es, Baldi:2011mt}).

\section{Conclusions}
\label{s:Con}

For a slowly rolling quintessence field, it is natural (though not necessary) that the field will eventually start oscillating at it approaches a minimum in its potential. In this paper we have analyzed the consequences on the expansion history and structure formation of a quintessence field, which initially behaves like dark energy and starts oscillating around the minimum in its potential at late times ($a\gtrsim0.8$). The potentials considered in detail here have a quadratic minimum and are shallower than quadratic away from the minimum. When a spatially homogeneous scalar field ($\vp$) oscillates about the minimum of such an anharmonic potential, it can pump energy into its spatially inhomogeneous perturbations ($\delta\vp$) through parametric resonance. The amount of energy transfer depends on both the wavenumber of $\delta\vp$ as well as the number of oscillations that take place in the background field. This leads to a rapid fragmentation of the homogeneous field and rapid resonant growth of the field fluctuations on time scales significantly shorter than $H_0^{-1}$.

In order to avoid discrepancies with the measured expansion history, and to simultaneously produce oscillations in the field, we have given a prescription for setting the initial conditions of the quintessence field and parameters in the potential. We have also explicitly shown that the potentials must be close to constant during the phase where the field is slowly rolling. Potentials that do not satisfy this requirement have too much of a delay between the end of slow roll and the beginning of oscillations, thus avoiding the rapid resonant growth of structure. Note that current data is consistent with a cosmological constant. Our model is in no way more natural than a cosmological constant or other dark energy models. However, we believe that the model's interesting phenomenology and potentially observable consequences warrants its study. 

Given a model that produces a background expansion history in good agreement with the measured expansion history, we have shown how the gravitational potential and the overdensity in WIMPs is affected by the resonant growth of the field fluctuations. We found that the metric perturbations develop scale-dependent growth, with the scale set by the mass of the scalar field potential $m\sim \sqrt{{U''(\vp\rightarrow 0)}}\sim 10^{3}H_0$ ($k\lesssim 0.1m$). On the other hand, the dark matter overdensity remains featureless and very similar to the \lcdm solution aside from an overall slight suppression because of the small difference in expansion history between \lcdm and the quintessence model. This is because, the dark matter does not have time to respond to the changing potential. Note that scale dependent changes in the gravitational potential are normally attributed to modified gravity (e.g. \cite{Amin:2007wi,Bertschinger:2008zb,Jain:2010ka}). Here, however, we have shown that such a change can occur in general relativity with a minimally coupled quintessence field with a canonical kinetic term. Thus, if future observations find evidence for scale-dependent growth, it cannot be attributed to modified gravity. 

The rapid growth of the potential significantly affects the ISW contribution to the temperature angular power spectrum of the CMB. For a range of parameters this yields the strongest constraint on such quintessence transitions. Since the metric perturbation develops a scale dependent change, the weak lensing power spectrum also offers a possible way to constrain this quintessence scenario, where dark energy undergoes a late-time transition described above. Unfortunately, because of the proximity of the decay in dark energy, deviations from \lcdm occur at low $l$ multipole moments. A full treatment, which includes nonlinearities in the quintessence field, however, could give rise to deviations in the weak lensing and temperature power spectrum, at larger $l$ values. 

The nonlinear dynamics of the quintessence field would give rise to a wealth of new phenomenon including field fragmentation and possible formation of localized scalar field lumps which could provide additional observational constraints. The full nonlinear analysis, combing N-body simulations for the dark matter and lattice simulations for the scalar field, is beyond the scope of this paper, but provides a promising avenue to explore quintessence transitions and their consequences further. Recall that in spite of its strong clustering properties, our quintessence field is minimally coupled with a canonical kinetic term and does not have nongravitational couplings to WIMP dark matter. This could make such models easier to simulate than models where gravity is modified (e.g. \cite{Laszlo:2007td,Schmidt:2009sg, Li:2011uw, Ferraro:2010gh, Zhao:2011cu}), interacting dark energy models  (e.g. \cite{Baldi:2008ay, Li:2010eu, Baldi:2010vv, Baldi:2011mt}), nonminimally coupled quintessence (e.g. \cite{RodriguezMeza:2009zz, Li:2010zw}) or when nonstandard kinetic terms are present.

In summary, if dark energy changes its nature close enough to the present time, it is possible to miss it in the expansion history measurements. For the models considered here, we have demonstrated that such a transition can dramatically change the gravitational potential power spectrum in a scale dependent way but leave the galaxy clustering unaffected, thus providing a possibly unique signature of such a transition. The best constraints on such transitions likely come from the ISW effect, followed by lensing. We expect, a more involved analysis, will provide additional constraints on such a transition once we include: (i) the nonlinear collapse in WIMPs and in the quintessence field, which will increase power on smaller scales; (ii) couplings to other fields (ignored here). It would also be interesting to explore similar resonant growth in models with multiple ultralight scalar fields (not necessarily dark energy) motivated in  \cite{Arvanitaki:2009fg} and studied in more detail in the context of structure formation by \cite{Marsh:2010wq}. 

\section{Acknowledgements}
We would like to thank Alan Guth, David Shirokoff, Raphael Flauger, Richard Easther, Andrew Liddle, Volker Springel, Robert Crittenden, Marco Bruni, Ignacy Sawicki, Jonathan Pritchard, Navin Sivanandam and Charles Shapiro for useful conversations. We would especially like to thank Mark Hertzberg for comments on an early draft of this work and Antony Lewis for repeated and prompt help when we were trying to check our ISW calculation using CAMB. We thank the anonymous referee for a number of constructive comments.
\section{Appendix A}
\subsection*{Floquet's theorem and calculating Floquet exponents}
The linearized equations of motion for the fluctuations, neglecting the Hubble expansion and metric perturbation, are (in Fourier space)
\Beq
&\partial_t^2\delta\varphi_k+\left[k^2+U''(\bp)\right]\delta\varphi_k=0,\\
\Eeq
where $k$ is the wavenumber. Since the homogeneous field $\vp$ is oscillating, $U''(\vp)$ is periodic in time. This results in a linear system with periodic coefficients that can be analyzed with Floquet theory. Floquet's theorem is most elegantly written in matrix form. Converting our second order equation of motion into a first order matrix equation, we find,
\Beq
\label{eq:dfq}
\partial_t x(t)=E(t)x(t),
\Eeq
where
\begin{displaymath}
x(t)=\left(\begin{array}{c}
\delta\varphi_k \\
\partial_t{\delta\varphi}_k
\end{array}  \right),
\end{displaymath}
and 
\begin{displaymath}
E(t)=\left(\begin{array}{cc}
0 & 1\\
-k^2-U''(\bp) & 0
\end{array}  \right).
\end{displaymath}
Before stating Floquet's theorem, we need one more definition. The fundamental matrix solution $\mathcal{O}(t,t_0)$, of Eq. \eqref{eq:dfq} satisfies
\Beq
\partial_t\mathcal{O}(t,t_0)&=E(t)\mathcal{O}(t,t_0),\\
\mathcal{O}(t_0,t_0)&=1.
\Eeq
The fundamental matrix solution evolves the initial conditions $x(t_0)$ in time: 
 \Beq
 x(t)=\mathcal{O}(t,t_0)x(t_0).
 \Eeq
 Explicitly, $\mathcal{O}(t,t_0)$ consists of two columns that represent two independent solutions $x_1, x_2$, which satisfy $x_1(t_0)=(1,0)$ and $x_2(t_0)=(0,1)$. Note that det $\mathcal{O}(t,t_0)$ is the Wronskian, det $\mathcal{O}(t_0,t_0)=1$, and there are no ``friction"  terms. Hence by Abel's Identity we have $\textrm{det}\, \mathcal{O}(t,t_0)=1.$ \\ \\ 
We are now ready to state {\it\bf Floquet's theorem} (without proof):\\ \\
Consider the linear system
\Beq
\partial_t x(t)=E(t)x(t),
\Eeq
where $x$ is a column vector and $E$ is a real, $2\times 2$ matrix satisfying $E(t+T)=E(t)$ for all $t$.
The fundamental solution $\mathcal{O}(t,t_0)$ is
 \Beq
\mathcal{O}(t,t_0)=P(t,t_0)\exp[(t-t_0)\mathcal{M}(t_0)],
\Eeq
where $P(t+T,t_0)=P(t,t_0)$ and $\mathcal{M}(t_0)$ satisfies $\mathcal{O}(t_0+T,t_0)=\exp[T \mathcal{M}(t_0)]$. The eigenvalues $\mu_{1,2}$ of $\mathcal{M}(t_0)$ are called Floquet exponents. 

Since det $\mathcal{O}(t_0+T,t_0)=1$, we have $\mu_1+\mu_2=0$. Suppose, from now on, that $\mu_1=-\mu_2=\mu$. 
If $\mathcal{M}(t_0)$ has two linearly independent eigenvectors $e_{\pm}(t_0)$ corresponding to $\pm\mu$ (including $\mu=0$), then the general solution can be written as
\Beq
x(t)=c_+\mathcal{P}_+(t,t_0) e^{{\mu}(t-t_0)}+c_-\mathcal{P}_-(t,t_0) e^{-\mu(t-t_0)}
\Eeq
where $\mathcal{P}_{\pm}(t,t_0)=P(t,t_0)e_{\pm}(t_0)$ are periodic column vectors. If there exists only one eigenvector corresponding to the repeated eigenvalue $\mu=0$, then the general solution becomes
\Beq
x(t)=c_+\mathcal{P}_+(t,t_0)+c_2\left[(t-t_0)\mathcal{P}_+(t,t_0) +\mathcal{P}_g(t,t_0)\right]
\Eeq
where $\mathcal{P}_{g}(t,t_0)=P(t,t_0)e_g(t_0)$ with $e_g(t_0)$ is a generalized eigenvector.

From above discussion, we have exponentially growing solutions iff the real part of the Floquet exponents $\Re[\mu]\ne0$. In order to calculate Floquet exponents $\pm\mu$, it is useful to consider the eigenvalues $\pi_{\pm}$ of $\mathcal{O}(t_0+T,t_0)$ called Floquet multipliers. Since det $\mathcal{O}(t_0+T,t_0)=1$, then $\pi_+\cdot\pi_-=1$. They are related to the Floquet exponents via $\pi_{\pm}=e^{\pm T\mu}$. In general $\mu$ and $\pi_\pm$ are complex. Using $\pi_{\pm}=|\pi_{\pm}|e^{i\theta_\pm}$ and $\pi_{+}\cdot\pi_-=1$, we have $$\mu=\frac{1}{T}\left[\ln |\pi_{+}|+i\theta_{+}\right]=-\frac{1}{T}\left[\ln |\pi_{-}|+i\theta_{-}\right].$$
Thus, $\Re[{\mu}]\ne 0$ if $|\pi_{\pm}|\ne 1$. 

Calculating the Floquet exponents explicitly for the problem at hand reduces to the following steps:
\begin{enumerate}
\item First we calculate the period $T$ of $U$. The period of $U(t)$ depends on the initial amplitude of the homogeneous field $\bp(t_0)$ (assuming $\partial_t\bp(t_0)=0)$ and is given by
\Beq
T(\bp_{max})=2\int_{\bp_{min}}^{\bp_{max}}\frac{d\bp}{\sqrt{2V(\bp_{max})-2V(\bp)}}.
\Eeq
In practice, we specify either $\bp_{max}$ or $\bp_{min}$. The other is found by solving $V(\bp_{min})=V(\bp_{max})$. For $V(\varphi)=V(-\varphi)$ we have $\bp_{max}=\bp_{min}$.
\item Next we solve $\partial_t \mathcal{O}(t,t_0)=E(t)\mathcal{O}(t,t_0)$ from $t_0$ to $t_0+T$ to obtain,
\begin{displaymath}
\mathcal{O}(t,t_0)=\left(\begin{array}{cc}
\delta\varphi^{(1)}_k(t_0+T) & \delta\varphi^{(2)}_k(t_0+T) \\
\partial_t\delta\varphi^{(1)}_k(t_0+T) & \partial_t\delta\varphi^{(2)}_k(t_0+T) 
\end{array}  \right)
\end{displaymath}
where $\{\delta\varphi^{(1)}_k(t_0)=1, \partial_t\delta\varphi^{(1)}_k(t_0)=0\}$, and $\{\delta\varphi^{(2)}_k(t_0)=0, \partial_t\delta\varphi^{(2)}_k(t_0)=1\}$. This is equivalent to solving $\partial_t^2\varphi_k+\left[k^2+V''(\bp)\right]\delta\varphi_k=0$ for the above two sets of initial conditions from $t_0$ to $t_0+T$. We have suppressed the dependence of $T$ on $\bp_{max}$ to reduce clutter.
\item Last, we find the eigenvalues of $\mathcal{O}(t_0+T,t_0)$. They are
\Beq
\pi_{\pm}=\frac{\delta\varphi^{(1)}_k +\partial_t\delta\varphi^{(2)}_k}{2}\pm\frac{\sqrt{\left\{\delta\varphi^{(1)}_k -\partial_t\delta\varphi^{(2)}_k\right\}^2+4\delta\varphi_k^{(2)}\partial_t\delta\varphi_k^{(1)}}}{2}
\Eeq
where all quantities are evaluated at $t_0+T$. Note that since the $\delta\varphi_k(t_0+T)$ depends on $k$, the eigenvalues also depend on $k$. The Floquet exponents $\pm\mu$ are then given by 
 \Beq
\pm\mu_k=\frac{1}{T}\ln\left[\frac{\delta\varphi^{(1)}_k +\partial_t\delta\varphi^{(2)}_k}{2}\pm\frac{\sqrt{\left\{\delta\varphi^{(1)}_k -\partial_t\delta\varphi^{(2)}_k\right\}^2+4\delta\varphi_k^{(2)}\partial_t\delta\varphi_k^{(1)}}}{2}\right].
\Eeq

\end{enumerate}

\section{Appendix B}
\subsection*{ISW effect from nonlinear dynamics of the quintessence field}
\noindent In the main body of the text, we discussed how the quintessence field fluctuations grow rapidly during the oscillatory regime. This rapid growth leads to a scale dependent growth in the gravitational potential which in turn leads to large changes in the CMB temperature anisotropies via the integrated Sachs-Wolfe effect (see Sec. 5.1). However, after $\anl\approx 0.95$ the field perturbations become nonlinear and the homogeneous field fragments rapidly, potentially forming long-lived, localized excitations of the field called oscillons. Our linear analysis did not include the ISW effect resulting from the nonlinear evolution of the field perturbations. In this Appendix, we remedy this by estimating the change in the CMB temperature due to the following:
\begin{itemize}
\item The change in the gravitational potential from the initial quasi/nonlinear evolution of the field perturbations.
\item The time varying gravitational potential due to an isolated oscillon after it is formed.
\end{itemize}
Let us first look at the ISW effect from the initial evolution of quasi/nonlinear perturbations. A realistic calculation requires input from detailed simulations of the nonlinear field dynamics (for example, \cite{Amin:2011hj}). Here we concentrate on a related toy problem of calculating the ISW effect due to the collapse of a single, spherical top-hat overdensity. This represents a crude approximation to the initial rapid quasi/nonlinear evolution of density perturbations associated with our model of quintessence. Later, we will evaluate the  CMB temperature anisotropies from the ISW effect due to a stable oscillon like configuration. Since the Universe expands very little between $\anl\approx 0.95$ and $a=1$, we will work in a Minkowski background. Our main purpose is to obtain an order of magnitude estimate and understand how the effect  depends on aspects of nonlinear evolution. 

Consider a spherical mass $\mathcal{M}$ with uniform density and a time dependent radius $R(t)$ located at $\bx_c$. The gravitational potential due to this mass is given by 

\[
  \Psi(t,\bx)= \left\{
  \begin{array}{l l}
    -\frac{G \mathcal{M}}{|\bx-\bx_c|}& \quad {|\bx-\bx_c|>R(t)},\\ \\
    \frac{G\mathcal{M}}{2R}\left(\frac{|\bx-\bx_c|^2}{R^2}-3\right)& \quad {|\bx-\bx_c|\le R(t)}.\\
  \end{array} \right.
\]
The partial time derivative of this potential is
\Beq
  \partial_t\Psi(t,\bx)=
    \frac{3G\mathcal{M}\dot{R}}{2R^4}u\Theta(u),
    \Eeq
    where $u=R^2(t)-|\bx-\bx_c|^2$ and $\Theta$ is the Heaviside function. The change in the CMB temperature due to a time varying potential (ISW term) is 
\Beq
\label{eq:ISWgen}
\frac{\Delta T(\bn)}{T}=2\int _{t_{\rm{nl}}}^{t_0}dt\,\,\partial_t\Psi(t,\bx)|_{\bx=(t_0-t)\bn},
\Eeq
where $\bn$ is a unit vector that defines observer's line of sight, $t_0$ is the time since last scattering $\approx 0.99H_0^{-1}$, and we replace $\bx$ by $(t_0-t)\bn$ since light travels on null geodesics. We have ignored anisotropic stress here ($\Psi=\Phi$). Note that the time of integration starts at $\tnl\approx 0.94H_0^{-1}$ (corresponding to $\anl\approx 0.95$) since nonlinear perturbations do not exist beforehand.

Let us consider ``linear" collapse, i.e.
\Beq
R(t)=\Rnl-v(t-\tnl),
\Eeq
where $0<v<1$. This yields
\Beq
u=(1-v^2)(t_+-t)(t-t_-),\quad t_{\pm}=\frac{t_c-v(\Rnl+v\tnl)\pm r_c}{1-v^2},
\Eeq
where
$t_c=t_0-\bx_c\cdot\bn=t_0-|\bx_c|\cos \theta$, $R_c=R(t_c)$, $r_c=\sqrt{R_c^2-(1-v^2)|\bx_c|^2\sin^2\theta}$ and $\theta$ is the angle between $\bn$ and $\bx_c$. Note that $t_{-}$ and $t_{+}$ denote the time a light ray enters and exist the spherical mass. With these definitions at hand, the integral over time in Eq. \eqref{eq:ISWgen} can be evaluated analytically. In particular,   if $\tnl<t_-<t_+<t_0$ (and $R_c>0$) we find
\Beq
\frac{\Delta T(\bn)}{T}=-\frac{4G\mathcal{M}v(1-v^2)^2r_c^3}{(R_c^2-v r_c^2)^2}.
\Eeq
Recalling that $r_c$ and $R_c$ depend on the angle $\theta$, the above expression shows that we should expect a cold-spot in the CMB, with an angular size $\sim R_c/|\bx_c|$. The anisotropy is negative and maximal when $\theta=0$ when the light ray passes through the center of the mass distribution,
\Beq
\frac{\Delta T(\bn)}{T}|_{\rm{max}}=-\frac{4G\mathcal{M}v}{R_c}.
\Eeq
The result depends only on  the size of the sphere $R_c$ when the light passes through its center, the speed of collapse $v$ and the mass $\mathcal{M}$. To obtain an order of magnitude estimate,
we use our Þducial set of parameters. We choose an initial perturbation of size $\Rnl\sim \knl^{-1}\approx (0.05m)^{-1}\approx0.02H_0^{-1}$ (see section 4.2.2 and 4.2.3), with density of order the critical density. We put the perturbation a distance  $|\bx_c|=0.03H_0^{-1}$ from us (corresponding $\sim (t_0-\tnl)/2$) and impose that it collapses with $v \sim 0.05$ (the approximate group velocity of a perturbation with wavenumber $\knl$). We get 
\Beq
\frac{\Delta T(\bn)}{T}|_{\rm{max}}\sim-4\times10^{-5},
\Eeq
which could provide additional constraints on our models. This signal is qualitatively similar to the Sunyaev-Zel'dovich temperature decrement from galaxy clusters \cite{Marriage:2010cp}. However unlike $\lesssim \rm{arcmin}$ angular scale of the SZ decrement, here, the angular scale is $\sim 30\, {\rm{degrees}}$. As can be checked, the conditions $\tnl<t_-<t_+<t_0$ and $R_c>0$ are satisfied for the above mentioned parameters. We remind the reader that this is merely an estimate based on a  spherical top-hat collapse. The actual nonlinear field evolution can be more complicated. In addition, the amplitude of the effect depends on the parameters in a nontrivial manner. Nevertheless, this exercise shows that it is indeed interesting to pursue the nonlinear evolution of the scalar field perturbations in detail. 

Let us now consider the ISW effect resulting from the time varying potential associated with a single oscillon. Based on our numerical solutions (radial only), oscillons with a central field amplitude of order $M$\footnote{This $M$ determines the scale where the scalar field potential changes shape, and should not be confused with $\mathcal{M}$ use to denote the mass of perturbations.} and width of order few $m^{-1}$, have a peak energy density of order $m^2M^2$ and an energy density profile that breathes in and out. This is in contrast with a fixed energy density profile often used in the oscillon literature. For an oscillon, crudely approximated by the a top-hat density configuration, we take  
\Beq
R_{\rm{osc}}(t) \approx 2m^{-1}\left(1+\frac{1}{4}\sin^2[m t+\phi]\right),
\Eeq
where for our fiducial model $m\sim 10^3H_0$. We stress that the actual energy density profile of oscillons in our model is more complicated. The energy density profile resembles a Gaussian when the central field amplitude is at its maximum, but becomes flatter when the field amplitude passes through zero. As a result, a calculation based on realistic oscillon profiles could be somewhat different from the estimates below. 

Unlike the linear collapse case, we did not find a simple analytic expression for the ISW contribution from individual oscillons. Evaluating it numerically, we find that for our fiducial set of parameters and assuming that the energy density within oscillons $\rho\sim m^2M^2\sim 3H_0^2\mpl^2$, we get 
\Beq
\frac{\Delta T(\bn)}{T}|_{\rm{max}}\sim\pm {G\mathcal{M}_{\rm{osc}}}m\times 10^{-2}\sim \pm10^{-7}.
\Eeq 
We have assumed that the oscillon is located at $|\bx_c|\sim 0.03H_0^{-1}$. The shape of the anisotropy pattern is ringlike. The phase of the radial oscillation $\phi$ determines whether the center has a temperature decrement or increment. Note that the radius  associated with an individual oscillon is small compared to the quasilinear perturbations  considered earlier in this Appendix (by an order of magnitude). As a result, one expects a smaller effect if one assumes that their density is still comparable to the average cosmological density. In addition, since the light crossing time is comparable to the oscillatory time scale of the energy density configuration, the ISW term undergoes cancellations, leading to a somewhat smaller ISW effect than what would be expected on purely dimensional grounds. Thus we expect the ISW contribution from individual oscillons to be smaller (and more localized) than that from the rapid collapse of nonlinear perturbations. 

Before we end this Appendix, we would like to comment on a caveat regarding the emergence of oscillons. Our linear analysis in the main body of the paper revealed that perturbations become nonlinear at $\anl$. While the nonlinear perturbations will form oscillons eventually, it is worth asking whether we can see them as large overdensities with individual identities by today. To unambiguously understand the timescales associated with emergence of oscillons would require a full lattice simulation which is beyond the scope of this current paper. Our analysis here indicates that it is certainly worth exploring this further. While here, we have concentrated on isolated inhomogeneities, it would be interesting to look at the combined effect of a collection of such inhomogeneities (with a number density of order $(\knl/2\pi)^3$) \cite{Amin:2010xe}, which would be closer to the actual scenario.



\end{document}